\documentclass[%
 reprint,
 amsmath,amssymb,
 aps,
pra
]{revtex4-1}
\usepackage{xcolor}
\usepackage{overpic}
\usepackage{hyperref}
\usepackage{dsfont}
\usepackage{bbold}
\usepackage{chbibref}
\usepackage{braket}
\usepackage{graphicx}
\usepackage{dcolumn}
\usepackage{bm}
\def\Xint#1{\mathchoice
   {\XXint\displaystyle\textstyle{#1}}%
   {\XXint\textstyle\scriptstyle{#1}}%
  {\XXint\scriptstyle\scriptscriptstyle{#1}}%
  {\XXint\scriptscriptstyle\scriptscriptstyle{#1}}%
  \!\int}
\def\XXint#1#2#3{{\setbox0=\hbox{$#1{#2#3}{\int}$}
     \vcenter{\hbox{$#2#3$}}\kern-.5\wd0}}

\def\dashint{\Xint-}

\begin{document}
\title{Open Quantum System Dynamics: \\ recovering positivity of the Redfield equation
via Partial-Secular Approximation} 

\author{Donato Farina}
\affiliation{Istituto Italiano di Tecnologia, Graphene Labs, Via Morego 30, I-16163 Genova, Italy} 
 \affiliation{NEST, Scuola Normale Superiore, I-56126 Pisa, Italy}
\email{donato.farina@sns.it}
\author{Vittorio Giovannetti}%
\affiliation{%
 NEST, Scuola Normale Superiore and Istituto Nanoscienze-CNR, I-56126 Pisa, Italy 
}%

\date{\today}
\begin{abstract}
We show how to recover complete positivity (and hence positivity) of the Redfield equation via a coarse grain average technique. We derive general bounds for the coarse graining time scale above which the positivity of the Redfield equation is guaranteed.
It turns out that a coarse grain time scale has strong impact on the characteristics of the Lamb shift term and implies in general non-commutation between the dissipating and the Hamiltonian components of the generator of the dynamical semi-group. 
Finally we specify the analysis to a two-level system or a quantum harmonic oscillator coupled to a fermionic or bosonic thermal environment via dipole-like interaction. 
\end{abstract}

\maketitle

\section{Introduction}
Describing in a proper way the whole time evolution of a quantum system interacting with a thermal environment is of crucial importance for quantum technology advancements and quantum computation \cite{nielsen2002quantum,riedel2017european}.
An exact treatment is in general a difficult task because of the exponential growth of the Hilbert space dimension with the number of constituents characterizing the system plus environment universe.
Reasonable assumptions (Born and Markov approximations) lead to the Redfield equation which while being quite effective in providing a 
reasonable description of several quantities of interest
in many contexts --- for example for excitons interacting with phonon baths with broad spectrum \cite{yang2002influence} --- 
 in general does not guarantee 
the positivity of the system evolution, nor the more stringent complete positivity condition, i.e. positivity of correlated states of the system with external degrees of freedom~\cite{nielsen2002quantum,HOLEVOBOOK}.
In practical terms this corresponds to the occurrence of non-positive eigenvalues in the system density matrix, resulting  in negative expectation probabilities for observables and, consequently, in unphysical predictions.
There are different ways of curing this non-positivity. For example in \cite{esposito2019infoflow, palmieri2009lindblad, kirvsanskas2018phenomenological} some reasonable assumptions on the bath correlation function are taken into account to handle this problem. Among these procedures the implementation of the secular approximation is a standard approach. 
It, on top of the Born and Markov approximations, results in the 
celebrated Gorini-Kossakowski-Lindblad-Sudarshan (GKLS) master equation (Lindblad equation) \cite{lindblad1976generators,gorini1976completely,breuer2002theory}.
However, when more than one characteristic frequency does characterize the internal dynamics of the system, the effectiveness of the secular approximation becomes an issue.
For instance its applicability is under current debate when the system is composed by two or more interacting parties, leading to the \textit{local vs global} discussion on Markovian master equations \cite{hofer2017markovian,gonzalez2017testing,de2018reconciliation}.  
In this context the global approximation makes full use of the secular approximation, leading to a steady state solution which is fully consistent with thermodynamics \cite{levy2014local}.
This is not the case
(even if a reconciliation can be achieved by taking into account the work produced by the switches of the environment ancillas in a collisional model approach
\cite{de2018reconciliation})
in general in a local approach \cite{levy2014local}, for which --- once assumed the smallness of the appropriate internal energy scales of the system --- the secular approximation is not required because the positivity of the system state turns out automatically.

In this paper we show how to recover complete positivity (and hence positivity) of the Redfield equation via a coarse grain average technique that is less drastic and have
less impact than the secular approximation.  
Contrarily to previous works \cite{lidar2001completely, majenz2013coarse, lidar2019lecture, stokes2012extending} we introduce a coarse grain time scale only once the Markov and Born approximation has been fully employed, allowing to study the effects of the secular approximation without introducing spurious contributions.
It turns out that this is feasible with a sufficiently large coarse grain time interval that depends on
 the spectrum of the system and on temperature.
An estimation of such critical  time scale  is derived under rather general assumptions on the system-bath coupling model. 
Furthermore we show that a finite coarse grain time has strong impact on the structure of the Lamb shift term and that typically it implies  non-commutation between the dissipating and the Hamiltonian parts of the generator of the resulting dynamical semi-group.

The sections are organized  as follows:
In Sec.\,\ref{sec:general} we introduce the general formalism of the partial secular approximation \cite{schaller2008preservation,schaller2009systematic,cresser2017coarse,cohen1998atom,seah2018refrigeration} when the Redfield equation is taken as starting point \cite{breuer2002theory}. We also discuss the above mentioned non-commutations between the Hamiltonian and dissipating components of the generator arising from the non-secular terms. 
Then we provide examples by specifying in Sec.\,\ref{sec:example-xx} the analysis to a qubit or a quantum harmonic oscillator system coupled to a fermionic or bosonic bath via dipole-like interaction.
Here the non secular terms come from the counter-rotating contributions of the coupling between the bath and the system degrees of freedom.
Finally in Sec.\,\ref{SEC.iv} we derive general bounds for the coarse graining time scale above which the positivity of the Redfield equation is guaranteed. \\
\section{\label{sec:general-formalism} From Redfield to GKSL via secular approximation}
\label{sec:general}
In this section we review the microscopic derivation of the Redfield equation 
and how one can arrive from it to a proper GKSL form via secular approximation.
 Following
Ref.~\cite{breuer2002theory} we shall work in a general setting,
limiting to a minimum all the assumptions on the system Hamiltonian  and on its 
environment. Readers who are familiar with the field can probably skip this part moving directly
to following sections  where our main results are presented.

\subsection{Microscopic Model}
\label{Sec-redfield}
Let S be a quantum system interacting with and external environment E. 
Following conventional approach we assume the SE compound to  be isolated and describe their joint evolution in terms of a global  Hamiltonian $H_{\rm SE}$  composed by  three terms:
\begin{eqnarray}
\label{eq:SE-Ham}
H_{\rm SE}=H_{\rm S}+H_{\rm E}+H_{\rm 1}\,,
\end{eqnarray}
with  $H_{\rm S}$ and $H_{\rm E}$ being local  contributions, and with $H_1$ being the coupling Hamiltonian which, in full generality, we express as 
\begin{eqnarray}
\label{eq:SE-interaction-Ham}
H_{1}=\sum_{\alpha=1}^M A_\alpha \otimes B_\alpha\,,
\end{eqnarray}
where $A_\alpha$ and $B_\alpha$ are not-null self-adjoint operators acting on S and E respectively, and where the parameter $M$ enumerates 
the number of non trivial terms entering the decomposition. 
As input state we take a factorized density matrix of the form
\begin{eqnarray} 
 \rho_{\rm SE}(0)=\rho_{\rm S}(0)\otimes \rho_{\rm E}(0)\;, \label{FACT} 
 \end{eqnarray} 
 with the  $\rho_{\rm E}(0)$ environment component   fulfilling  the following {\em stationary} conditions: 
 \begin{itemize} 
\item invariance under the action of the local Hamiltonian, i.e. 
 \begin{eqnarray} 
 \Big[\rho_{\rm E}(0),H_{\rm E}\Big]_-=0 \;;  \label{S1} 
 \end{eqnarray} 
 where hereafter the symbols $[\cdots, \cdots]_{\pm}$ will be used to represent the commutator ($-$) and
 the anti-commutator ($+$), respectively; 
\item   zero   expectation value of the operators $B_\alpha$ entering the coupling Hamiltonian~(\ref{eq:SE-interaction-Ham}), i.e.
  \begin{eqnarray} 
\mbox{Tr}_{\rm E} \{  \rho_{\rm E} (0) B_\alpha \} =0   \;, \qquad \forall \alpha\in\{ 1, \cdots, M\} \;, \label{S2} 
 \end{eqnarray} 
the symbol $\mbox{Tr}_{\rm E} \{ \cdots \}$ representing the partial trace with respect to the environment degrees of freedom.  
 \end{itemize} 
 As we shall see in the following the condition~(\ref{S2}) is essential for dropping first order contributions
 in the system master equation: it should be stressed that however it is not as stringent as it may looks at first site, as it  can always be enforced by properly redefining the free Hamiltonian of $S$.

We hence move in  the interaction picture in which the free Hamiltonian of the universe $H_0=H_{\rm S}+H_{\rm E}$
is integrated away,  introducing the operators
 \begin{eqnarray}
\label{eq:universe-operator-int-pict}
\tilde{H}_1(t)&:=&e^{i H_{\rm 0} t} {H}_1e^{-i H_{\rm 0} t}\,,  \\ 
\tilde{\rho}_{\rm SE}(t)&:=&e^{i H_{\rm 0} t} \rho_{\rm SE}(t) e^{-i H_{\rm 0} t}\,,\\
\tilde{\rho}_{\rm S}(t)&:=&\mbox{Tr}_{\rm E} \{ \tilde{\rho}_{\rm SE}(t) \} = 
e^{i H_{\rm S} t} \rho_{\rm S}(t) e^{-i H_{\rm S} t}\,,
\end{eqnarray} 
with $\rho_{\rm SE}(t)$ the density matrix of SE  at time $t$ and
 $\rho_S(t):=
\mbox{Tr}_{\rm E} \{ {\rho}_{\rm SE}(t) \}$ its reduced form describing the corresponding state of S
($\hbar$ having been set equal to 1). 
Accordingly the dynamics of the joint system writes
$\dot{\tilde{\rho}}_{\rm SE}(t)=- i [\tilde{H}_1(t), \tilde{\rho}_{\rm SE}(t)]_-$ which, 
upon formal integration,  
can be equivalently expressed as 
\begin{eqnarray}
\dot{\tilde{\rho}}_{\rm SE}(t)
&=&  - i \Big[\tilde{H}_1(t), \rho_{\rm SE}(0)\Big]_- \label{eq:universe-eq-motion-int-pict} 
 \\ 
&& - \int_0^t d\tau \Big[ \tilde{H}_1(t), \Big[\tilde{H}_1(t-\tau), \tilde{\rho}_{\rm SE}(t-\tau)\Big]_-\Big]_- \;.
\nonumber
\end{eqnarray}
Taking the partial trace with respect to E the left-hand-side of Eq.~(\ref{eq:universe-eq-motion-int-pict}) 
reduces to  the first derivative of $\tilde{\rho}_{\rm S}(t)$ while the
 first term on the right-hand-side disappears thanks to the cooperative effect of the stationary conditions (\ref{S1}) and (\ref{S2}).
 The integral contribution on the contrary still exhibits a non-trivial  functional dependence on the joint state  $\tilde{\rho}_{\rm SE}(t)$ which we
 treat by invoking the {\it Born} (or {\it weak-coupling})  {\it approximation}, 
 requiring that at first order in the coupling  the state of the environment is not affected by the presence of S, i.e. writing 
\begin{eqnarray} 
\tilde{\rho}_{\rm SE}(t_1)\simeq \tilde{\rho}_{\rm S}(t_1)\otimes \tilde{\rho}_{\rm E}(0)\,,\end{eqnarray} 
all $t_1\in [0,t[$. 
Under this condition we hence arrive to the following homogenous equation for S, 
\begin{eqnarray}
\label{eq:born}
\dot{\tilde{\rho}}_{\rm S}(t)&\simeq& 
\int_0^t d\tau \sum_{\alpha, \beta=1}^M c_{\alpha \beta}(\tau) 
\Big( \tilde{A}_{\beta}(t-\tau) \tilde{\rho}_{\rm S}(t-\tau) \tilde{A}_{\alpha}(t)\nonumber\\
&-& \tilde{A}_{\alpha}(t) \tilde{A}_{\beta}(t-\tau) \tilde{\rho}_{\rm S}(t-\tau)\Big) + h.c.\,,
\end{eqnarray} 
with $\tilde{A}_\alpha(t) = e^{i H_{\rm S} t} {A}_\alpha e^{-i H_{\rm S} t}$ and 
where $c_{\alpha \beta}(\tau)$ are 
environment correlation functions  defined as
\begin{eqnarray}
c_{\alpha \beta}(\tau):={\rm Tr_{E}} \{\rho_{\rm E}(0) e^{i H_{\rm E} \tau} {B}_\alpha e^{-i H_{\rm E} \tau} B_\beta\}\,,
\label{eq:bath-correlations}
\end{eqnarray} 
that exploiting  Eq.~(\ref{S1}) and the fact that 
the $B_\alpha$s  are self-adjoint operators  can be shown to 
fulfil  the condition
\begin{eqnarray} \label{SYM} 
c^*_{\alpha \beta}(\tau) = c_{\beta \alpha}(-\tau) \;. \end{eqnarray} 
The next assumption concerns the memory properties of the environment.
We call $\tau_{\rm E}$ the characteristic width of the environment correlation functions $c_{\alpha \beta}(\tau)$
and we assume that the time scales $\delta t$ over which the system S significantly evolves in the interaction picture
satisfy the condition $\delta t\gg \tau_{\rm E}\,.$ 
This hypothesis justifies 
the \textit{Markov approximation} which in Eq.~(\ref{eq:born}) neglects \textit{i)} the $\tau$ dependence of the state and \textit{ii)} substitutes the upper extreme of integration with $+\infty$, leading to the Redfield equation \cite{redfield1957theory,breuer2002theory}
\begin{eqnarray}
\dot{\tilde{\rho}}_{\rm S}(t)&\simeq&
\int_0^\infty d\tau  \sum_{\alpha, \beta=1}^M c_{\alpha \beta}(\tau) 
\Big(  \tilde{A}_\beta(t-\tau) \tilde{\rho}_{\rm S}(t) \tilde{A}_\alpha(t)\nonumber \\ 
 &-& \tilde{A}_\alpha(t)  \tilde{A}_\beta(t-\tau) \tilde{\rho}_{\rm S}(t) \Big) + h.c.
 \label{eq:redfield-generale} \\
 &=& \sum_{i j}
{\Gamma}_{ij}(t)
\left(A_{j}^\dagger \tilde{\rho}_{\rm S}(t) A_{i} - A_{i} A_{j}^\dagger  \tilde{\rho}_{\rm S}(t)\right)
+ h.c.\, , \nonumber 
\end{eqnarray} 
where the last identity has being obtained by decomposing the 
operators $A_\alpha$ in terms of the eigenvectors 
 of the free system Hamiltonian.
 Specifically we write 
\begin{eqnarray}
A_\alpha=\sum_\omega A_{\alpha \omega}\;,
\end{eqnarray} 
with
\begin{eqnarray}
A_{\alpha \omega}&:=&\sum_{\epsilon_1\,,\, \epsilon_2: \epsilon_1-\epsilon_2=\omega}
\pi_{\epsilon_1} A_\alpha \pi_{\epsilon_2}=
\sum_{\epsilon}
\pi_{\epsilon+\omega} A_\alpha \pi_{\epsilon}
\,,
\label{eq:A-eigenstates}
\end{eqnarray} 
where $\pi_\epsilon$ is the projector associated with the eigenvalue $\epsilon$ of $H_{\rm S}$,  i.e. $H_{\rm S}=\sum_\epsilon \epsilon \; \pi_\epsilon$. 
The new variable 
$\omega:=\epsilon_1-\epsilon_2$ spans a range of $G$ different cases, counting
 all the distinguishable energy gaps of $H_{\rm S}$ (including the zero energy gap value
 associated with the terms where $\epsilon_1=\epsilon_2$). 
Introducing then the collective indices $i=(\alpha, \omega)$ and $j=(\beta, {\omega'})$ 
which run over a set of $N=GM$ different entries, 
and noticing that $A_{\beta -{\omega'}}=A^\dagger_{\beta {\omega'}}$, the Redfield equation~(\ref{eq:redfield-generale}) can hence be casted as shown in the last line with the $N\times N$ matrix  ${\Gamma}_{ij}(t)$ given by 
\begin{eqnarray}
{\Gamma}_{ij}(t)=e^{i(\omega-{\omega'})t} \Omega_{\alpha \beta}({\omega'})\;,
\end{eqnarray} 
where for each value of the energy gap $\omega$ the coefficients 
\begin{eqnarray} \label{MATRIXOMEGA} 
 \Omega_{\alpha \beta}(\omega) &:=& \int_0^\infty d\tau c_{\alpha \beta}(\tau) e^{i \omega \tau}\;,
\end{eqnarray} 
identify an   $M\times M$ complex matrix $\Omega(\omega)$ that is going to play an important role in what follows.
Equation~(\ref{eq:redfield-generale})  can be further simplified by performing 
a temporal averaging over coarse grain time intervals $\Delta t$ which is larger than $\tau_{\rm E}$ but sill much smaller than the 
time scale $\delta t$ where
$\tilde{\rho}(t)$  varies appreciably, i.e. 
\begin{eqnarray} \tau_{\rm E} \ll \Delta t\ll \delta t \;. \label{COARSELIMIT} \end{eqnarray} 
The last passage is intimately connected with the hypothesis underlying the Markov approximation and, as we shall see in the following, is essential in order to recover the 
GKSL structure of the generator.  In particular using the fact that the coarse graining does not affect 
$\tilde{\rho}(t)$,  we can replace  (\ref{eq:redfield-generale}) with 
\begin{equation}
\label{eq:general-non-can-me-int-pict}
\dot{\tilde{\rho}}_{\rm S}(t)\simeq \sum_{i j}
{\Gamma}_{ij}^{(\Delta t)}(t)
\left(A_{j}^\dagger \tilde{\rho}_{\rm S}(t) A_{i} - A_{i} A_{j}^\dagger  \tilde{\rho}_{\rm S}(t)\right)
+ h.c.\, ,
\end{equation}
where now 
\begin{equation} \label{COARSE} 
{\Gamma}_{ij}^{(\Delta t)}(t): =\frac{1}{\Delta t} \int_{t-\Delta t/2}^{t+\Delta t/2} ds\; {\Gamma}_{ij}(s) 
= {\Gamma}_{ij}(t)\;S_{\omega - {\omega'}}^{(\Delta t)}\;, 
 \end{equation} 
where we introduced the function 
\begin{eqnarray}
\label{eq:S-function-def}
S_{\omega - \omega'}^{(\Delta t)}&:=& \mbox{sinc}[(\omega-\omega') \Delta t/2] \;, 
\end{eqnarray}
with $\mbox{sinc}[x]: = {\sin x}/{x} $ being the cardinal sinus. 

Next we express the matrix $\tilde{\Gamma}_{ij}^{(\Delta t)}(t)$  in terms of its hermitian and anti-hermitian components 
 writing 
 \begin{eqnarray}
 \label{eq:general-gamma-psa}
{\Gamma}_{ij}^{(\Delta t)}(t) = {\gamma}_{ij}^{(\Delta t)} (t)/2+i\; 
{\eta}_{ij}^{(\Delta t)}(t)\;, 
\end{eqnarray} 
with 
\begin{eqnarray}
\label{eq:hermitian-part-of-gammaij}
{\gamma}_{ij}^{(\Delta t)}(t)&:=&{\Gamma}_{ij}^{(\Delta t)}(t) + ({\Gamma}^{(\Delta t)}_{ji}(t))^*\,, \\
\label{eq:anti-hermitian-part-of-gammaij}
{\eta}_{ij}^{(\Delta t)}(t) &: =& \Big( {\Gamma}_{ij}^{(\Delta t)}(t)-({\Gamma}^{(\Delta t)}_{ji}(t))^*\Big)/(2 i)\,.
\end{eqnarray}  
With this choice the terms on the r.h.s. of Eq.~(\ref{eq:redfield-generale})  can be expressed as 
  \begin{eqnarray}
\label{eq:me-general-int-pict}
\dot{\tilde{\rho}}_{\rm S}(t)&\simeq &-i\left[\tilde{H}^{(\Delta t)}_{\rm LS}(t), \tilde{\rho}_{\rm S}(t)\right]_-\\
&&+\sum_{ij}{\gamma}_{ij}^{(\Delta t)}(t) \Big(A_j^\dagger \tilde{\rho}_{\rm S}(t) A_i 
- \frac{1}{2} \Big[ A_i A_j^\dagger, \tilde{\rho}_{\rm S}(t)\Big]_+\Big)\,, \nonumber
\end{eqnarray} 
where $\{ \cdots, \cdots\}$ in the second line represents the anti-commutator and
$\tilde{H}_{\rm LS}^{(\Delta t)} (t)$  the Lamb shift term 
\begin{eqnarray}
\label{eq:lamb-int-pict}
\tilde{H}_{\rm LS}^{(\Delta t)}(t):=\sum_{ij}  {\eta}_{ij}^{(\Delta t)}(t)  \; A_i A_j^\dagger\,.
\end{eqnarray} 
Going back in  ${\rm Schr\ddot{o}dinger}$ picture we can finally remove the time dependence of coefficients $\gamma_{ij}^{(\Delta t)}(t)$ and  $\eta_{ij}^{(\Delta t)}(t)$  obtaining  a master equation with constant generator terms 
\begin{eqnarray}
\label{eq:me-general-Sch-pict}
\dot{\rho}_{\rm S}(t)&\simeq &-i\left[H^{(\Delta t)}_{\rm S}, \rho_{\rm S}(t)\right]_-\\
&& +\sum_{ij}\gamma^{(\Delta t)}_{ij}
\Big(A_j^\dagger \rho_{\rm S}(t) A_i 
- \frac{1}{2} \Big[A_i A_j^\dagger, \rho_{\rm S}(t)\Big]_+\Big)\,, \nonumber
\end{eqnarray} 
where now 
\begin{eqnarray} 
H^{(\Delta t)}_{\rm S}&:=& H^{(\Delta t)}_{\rm LS}+H_{\rm S} \;,\label{HAMDT}  \\
\label{HLSsc} 
H_{\rm LS}^{(\Delta t)}&:=&\tilde{H}_{\rm LS}^{(\Delta t)}(0)=  \sum_{ij} \eta_{ij}^{(\Delta t)} A_i A_j^\dagger \;.\end{eqnarray} 
Explicitly, exploiting the symmetry~(\ref{SYM}), the  $N\times N$
 matrices 
$\gamma^{(\Delta t)}_{ij}$
and $\eta_{ij}^{(\Delta t)}$ appearing in these expressions can be shown to correspond to
\begin{eqnarray} \label{DEFGAMMA} 
 \gamma^{(\Delta t)}_{ij}&:=& \gamma^{(\Delta t)}_{ij}(0) = 
   \gamma^{(+)}_{\alpha \omega,\beta {\omega'}} 
\; S_{\omega - {\omega'}}^{(\Delta t)}\,, \\
\eta^{(\Delta t)}_{ij}&:=&\eta_{ij}^{(\Delta t)}(0) = 
 \frac{ \gamma^{(-)}_{\alpha \omega,\beta {\omega'}}}{2i} 
\; S_{\omega - {\omega'}}^{(\Delta t)}\,, \label{DEFETA} 
\end{eqnarray} 
with  
\begin{eqnarray}
\label{gamma-pm-def}
 \gamma^{(\pm)}_{\alpha \omega,\beta {\omega'}} &:=& 
 \Omega_{\alpha \beta}({\omega'})\pm \Omega_{\beta \alpha}^*(\omega) \;.
\end{eqnarray}

The last passage needed to put Eq.~(\ref{eq:me-general-Sch-pict}) in GKSL is the  diagonalization of 
$\gamma_{ij}^{(\Delta t)}$. Such step however works if and only if such matrix is positive semi-definite (or equivalently non-negative), the presence of negative eigenvalues being formally incompatible with the complete-positivity requirement~\cite{HOLEVOBOOK} of 
 the resulting dynamics of $\rho_{\rm S}(t)$.
 This is the reason  for which one introduces the coarse-graining transformation~(\ref{COARSE}).
 Indeed thanks to the fact that 
 \begin{eqnarray}
\lim_{\Delta t\rightarrow \infty} S_{\omega - {\omega'}}^{ (\Delta t)} = \delta_{\omega,{\omega'}} \;,
\end{eqnarray} 
  as  $\Delta t$ diverges the $N\times N$ matrix 
$\gamma^{(\Delta t)}_{i j}$ 
reduces to a block diagonal form with respect to the frequency labels, 
\begin{eqnarray} \label{DEFGAMMAinfy} 
\gamma^{(\infty)}_{ij}&:=&\lim_{\Delta t\rightarrow \infty} \gamma^{(\Delta t)}_{ij}  = \gamma^{(+)}_{\alpha \omega,\beta \omega}  
\delta_{\omega,\omega'} \;, 
\end{eqnarray} 
 where for each $\omega$ the coefficients  $\gamma^{(+)}_{\alpha \omega,\beta \omega}$ identify $M\times M$ matrices
 \begin{eqnarray} \label{DEFGAMMAOMEGAOMEGA}
 \gamma^{(+)}(\omega,\omega):= \Omega(\omega) +  \Omega^\dag(\omega)\;,\end{eqnarray} 
    that are 
explicitly 
non-negative
(see Appendix~\ref{appA} for details). 
The  $\Delta t\rightarrow \infty$ limit goes under 
the name of {\it secular approximation} (SA) and it is the last  step one typically enforces in order to 
recover the GKSL form~\cite{breuer2002theory}: 
this is a rather drastic approximation, which, formally speaking, is an explicit violation of the 
 upper bound~(\ref{COARSELIMIT}) and which forces  two main structural constraints on the resulting
master equation (ME). Specifically from~(\ref{DEFETA}) it follows that under SA also the matrix $\eta^{(\Delta t)}_{ij}$ gets
block diagonal with respect to the gap indexes $\omega$ and $\omega'$, 
\begin{eqnarray} \label{DEFetainfy} 
\eta^{(\infty)}_{ij}&:=&\lim_{\Delta t\rightarrow \infty} \eta^{(\Delta t)}_{ij}  = \frac{\gamma^{(-)}_{\alpha \omega,\beta \omega} }{2i}  
\; \delta_{\omega,\omega'} \;, 
\end{eqnarray} 
yielding the following properties
\begin{itemize} 
\item[i)] 
 commutation between 
the Lamb shift Hamiltonian $H_{\rm LS}^{(\infty)}$  and the free Hamiltonian contribution  $H_{\rm S}$, (see e.g.
Eq.~(\ref{THISone}) below);  
\item[ii)] commutation between the free Hamiltonian $\mathcal{H}_{\rm S}$  and the dissipative super-operator components $\mathcal{D}^{(\infty)}$ of  the 
dynamical semi-group generator;
\item[iii)] under certain hypotheses, commutation between the full Hamiltonian $\mathcal{H}_{\rm S}^{(\infty)}$ super-operator and $\mathcal{D}^{(\infty)}$.
\end{itemize} 
The first property can be easily verified by  
expanding 
  the indexes $i$, $j$ appearing in Eq.~(\ref{HLSsc}) and using the property
  \begin{eqnarray} 
   H_{\rm S} \pi_\epsilon =  \pi_\epsilon H_{\rm S}=  \epsilon  \pi_\epsilon\;.
\end{eqnarray} 
Accordingly we get 
\begin{eqnarray} \label{COM1} 
\left[H_{\rm S},H^{(\Delta t)}_{\rm LS}\right]_-&=&
\sum_{\alpha \beta \omega {\omega'}}
(\omega-{\omega'})  \; \eta^{(\Delta t)}_{\alpha \omega, \beta {\omega'}} \nonumber \\
&\times& 
 \sum_{\epsilon} \pi_{\epsilon+\omega} A_\alpha \pi_{\epsilon} A_\beta 
 \pi_{\epsilon + {\omega'}}\,, \end{eqnarray} 
which in the SA limit  where  Eq.~(\ref{DEFetainfy}) forces $\eta^{(\Delta t)}_{\alpha \omega, \beta {\omega'}}$ to be proportional to the Kronecker delta $\delta_{\omega,\omega'}$, gets explicitly null, i.e. 
\begin{eqnarray} 
 \lim_{\Delta  t \rightarrow \infty} \left[H_{\rm S},H^{(\Delta t)}_{\rm LS}\right]_- = 
 \left[H_{\rm S},H^{(\infty)}_{\rm LS}\right]_- =0 \;. \label{THISone} 
 \end{eqnarray} 
To properly express property ii)  let us rewrite the r.h.s. of Eq.~(\ref{eq:me-general-Sch-pict}) in the formal
compact way:
\begin{eqnarray} 
\label{eq:generator-delta-t}
 {\cal L}^{(\Delta t)} [{\rho}_{\rm S}(t)] := {\cal H}^{(\Delta t)}[{\rho}_{\rm S}(t)]
+ {\cal D}^{(\Delta t)}[{\rho}_{\rm S}(t)]\;,
\end{eqnarray} 
where ${\cal H}_{\rm S}^{(\Delta t)}:= {\cal H}_{\rm S} + {\cal H}^{(\Delta t)}_{\rm LS}$ and ${\cal D}^{(\Delta t)}$ represent the Hamiltonian and  dissipative contributions
to the super-operator ${\cal L}^{(\Delta t)}$ which generates  the system dynamics, i.e. 
\begin{eqnarray} 
{\cal H}_{\rm S}[ \cdots ]&:=& -i \Big[ H_{\rm S} ,\cdots \Big]_- \;, \\
{\cal H}_{\rm LS}^{(\Delta t)}[ \cdots ]&:=& -i \Big[ H^{(\Delta t)}_{\rm LS} ,\cdots \Big]_- \;, \\
{\cal D}^{(\Delta t)}[ \cdots ]&:=& \sum_{ij}\gamma^{(\Delta t)}_{ij}
\Big(A_j^\dagger \cdots A_i 
- \frac{1}{2} \Big[ A_i A_j^\dagger, \cdots \Big]_+\Big) \;. \nonumber \\
\end{eqnarray} 
The commutation between $\mathcal{H}_{\rm S}$ and $\mathcal{D}^{\rm (\infty)}$, i.e. 
\begin{equation}
\label{Hs-D-commute}
\left[\mathcal{H}_{\rm S}, \mathcal{D}^{\rm (\infty)}\right]_-:=
\mathcal{H}_{\rm S}\circ \mathcal{D}^{\rm (\infty)}-
\mathcal{D}^{\rm (\infty)} \circ \mathcal{H}_{\rm S}
=0\;,
\end{equation}
with "$\circ$" being the composition of super-operators, 
can then be proven by inspection, exploiting that, by construction, the operators $A_{\alpha \omega}$ are eigen-operators of
$H_{\rm S}\,,$ i.e.
\begin{equation}
\label{eigen-operators}
\left[H_{\rm S}, A_{\alpha \omega}\right]_-=\omega A_{\alpha \omega}\,.
\end{equation}
We discuss now the point iii).
Despite Eqs.~(\ref{THISone}) and (\ref{Hs-D-commute}) the generators $\mathcal{H}_{\rm LS}^{(\infty)}$
and 
$\mathcal{D}^{(\infty)}$
in general don't commute
(and consequently  
$\mathcal{H}_{\rm S}^{(\infty)}$
and 
$\mathcal{D}^{(\infty)}$
neither)
if we don't enforce some specific hypotheses.
Similarly to the property in Eq.~(\ref{eigen-operators}), a sufficient condition for the commutator
$ \left[\mathcal{H}^{\rm (\infty)}_{\rm LS}, 
\mathcal{D}^{\rm (\infty)}\right]_-$
to be zero is  to have the operators $A_{\alpha \omega}$  eigen-operators of $H_{\rm LS}^{(\infty)}$
with eigenvalues $f(\omega)\,,$ the last being an odd function of $\omega$, i.e. 
\begin{equation}
\label{labshift-eigenoperator}
\left[H_{\rm LS}^{(\infty)}, A_{\alpha \omega}\right]_-=f(\omega) A_{\alpha \omega}\,, \qquad 
f(-\omega)=-f(\omega)\,.
\end{equation} 
This is verified for instance when \textit{(a)} 
both the gaps $\omega$ and the energies $\epsilon$ are non-degenerate, i.e. for a given energy gap $\omega$ we associate one and only one pair $(\epsilon_1, \epsilon_2)$, with $\epsilon_i$ non-degenerate eigenvalues of $H_{\rm S}$ or \textit{(b)} when the energy levels are of the type $\epsilon_n=n \omega_0$ --- implying that the energies are non-degenerate, but the gaps are --- and the only effect of $H_{\rm LS}^{(\infty)}$ is a renormalization of the characteristic energy $\omega_0\,.$ Examples \textit{(a)} and \textit{(b)} will be presented in the next Section. 
Here we just notice that when condition in Eq.~(\ref{labshift-eigenoperator}) holds, because of Eq.~(\ref{Hs-D-commute}),
we achieve commutation also between the Hamiltonian and the dissipator:
\begin{equation}
\label{H-D-commute}
\left[\mathcal{H}_{\rm S}^{\rm (\infty)}, 
\mathcal{D}^{\rm (\infty)}\right]_-=0\,.
\end{equation}
Remarkably, as we shall see explicitly in the next Section, going beyond the SA by working with finite values of 
the coarse graining time $\Delta t$, in general one has 
\begin{eqnarray}
\label{Hs-Hls-commutator-PSA}
\left[H_{\rm S}, H_{\rm LS}^{(\Delta t)}\right]_-\neq 0\,,
\end{eqnarray}
and
\begin{eqnarray}
\label{Hs-D-commutator-PSA}
\left[\mathcal{H}_{\rm S}, \mathcal{D}^{(\Delta t)}\right]_-\neq 0\,.
\end{eqnarray}
Furthermore when Eq.~(\ref{H-D-commute}) is satisfied within SA, 
 the breaking of commutation rules in Eqs.~(\ref{Hs-Hls-commutator-PSA}), (\ref{Hs-D-commutator-PSA}) can induce non-commutation also between 
$\mathcal{H}_{\rm S}^{(\Delta t)}$ and 
$\mathcal{D}^{(\Delta t)}$, i.e. 
\begin{eqnarray}
\label{H-D-commutator-PSA}
\left[\mathcal{H}_{\rm S}^{(\Delta t)}, \mathcal{D}^{(\Delta t)}\right]_-\neq 0\,.
\end{eqnarray}
\section{Partial Secular approximation}\label{sec:partial} 
 While effective in transforming  the Redfield equation into a GKSL dynamical semigroup, hence restoring the complete-positivity of the
 resulting dynamics, the SA is not strictly necessary. As a matter of fact in many models of physical interest, it is possible to arrive to a proper GKSL form also by adopting a Partial Secular Approach (PSA) where the 
 coarse graining step  is performed over time scales  $\Delta t$ which are finite. 
Explicit examples will be presented in this section, while in 
Sec.~\ref{SEC.iv} a set of sufficient conditions that allows one to determine the range of such special coarse graining times, will be given 
in a rather general context.
As we shall see under PSA,  while the complete-positivity of the Redfield equation is maintained, 
three main structural modifications can occur, namely the loss of the commutation relations 
i), ii) and iii) 
detailed in the previous section.

\subsection{\label{sec:example-xx}  An application to qubit and harmonic oscillator models}

The method of the PSA can be applied in the case of a single qubit or quantum harmonic oscillator (QHO)  coupled to a fermionic or bosonic bath via dipole-like interaction. 
As in Eq.~(\ref{eq:SE-Ham}), the Hamiltonian of the total system $H_{\rm SE}=H_{\rm S}+ H_{\rm E} + H_{1}$ is composed of three terms with
\begin{eqnarray}
\label{eq:xx-dipole-1}
H_{\rm S}&=&\omega_0 \zeta^\dagger \zeta\,,\\
\label{eq:xx-dipole-2}
H_{\rm E}&=&\sum_k \omega_k c_k^\dagger c_k\,,\\
\label{eq:xx-dipole-3}
H_{\rm 1}&=&\sum_k \gamma_k (c_k^\dagger + c_k)(\zeta+\zeta^\dagger)\,.
\end{eqnarray} 
Equations (\ref{eq:xx-dipole-1}) and (\ref{eq:xx-dipole-2}) describe the free Hamiltonians of the system and of the environment respectively and Eq.~(\ref{eq:xx-dipole-3}) is the system-environment interaction.
The ladder operators of the system $\zeta$ and $\zeta^\dagger$ and the ones of the environment $c_k$ and $c_k^\dagger$ respect the following commutation rules:
\begin{eqnarray}
\zeta \zeta^\dagger - s \zeta^\dagger \zeta&=&1\\
c_k c_{k'}^\dagger - q c_{k'}^\dagger c_k&=&\delta_{k, k'}\\
c_k c_{k'} - q c_{k'} c_k&=&0\,.
\end{eqnarray} 
In this formalism
$s=1\,(-1)$ implies that the system S is a QHO (qubit) and $q=1\,(-1)$ implies that the environment E is a bosonic (fermionic) thermal bath.
Under these assumptions the Redfield Eq.~(\ref{eq:redfield-generale})
reduces to:
\begin{eqnarray}
\label{eq:redfield-xx}
\dot{\tilde{\rho}}_{\rm S}(t)= \int_0^\infty d\tau c(\tau) &\times&  \\
\{\tilde{A}(t-\tau) \tilde{\rho}_{\rm S}(t) \tilde{A}(t) &-&\tilde{A}(t)  \tilde{A}(t-\tau) \tilde{\rho}_{\rm S}(t)\} + h.c.\,. \nonumber
\end{eqnarray} 
Here we have no index $\alpha$ since the interaction in Eq.~(\ref{eq:xx-dipole-3}) is a single tensor product (M=1) of two hermitian operators A and B, the first on the system and the second on the bath: $A=\zeta+\zeta^\dagger$, $B=\sum_k \gamma_k (c_k + c_k^\dagger)$. 
This leads to a single bath correlation function $c(\tau)$ (see Eq.~(\ref{eq:bath-correlations})) which is convenient to split into two terms:
\begin{equation}
\label{eq:bath-corr-func-xx}
c(\tau):=\left< \tilde{B}(\tau) B \right> = c_1(\tau)+c_2(\tau)\,,
\end{equation} 
with
\begin{eqnarray}
\label{eq:bath-corr-func-xx_abs}
c_1(\tau)&:=&\sum_k \gamma_k^2 n_k e^{i \omega_k \tau}\,,
\\
\label{eq:bath-corr-func-xx_em}
c_2(\tau)&:=&\sum_k \gamma_k^2 (q n_k + 1) e^{-i \omega_k \tau}\,,
\end{eqnarray} 
with $n_k:=\langle c_k^\dagger c_k \rangle$ being the occupation number at wave vector $k\,,$ following the Bose-Einstein (Fermi-Dirac) 
distribution
for $q=1$ $(q=-1)\,:$
\begin{eqnarray}
n_k=\frac{1}{e^{\beta \omega_k}- q}\,.
\end{eqnarray} 
Notice that the value of $q$ determines also the sign of the term into $c_2(\tau)$ (see Eq.~(\ref{eq:bath-corr-func-xx_em})), which is responsible for the stimulated emission in the standard SA. 
The expression of the system operator A in interaction picture,
$\tilde{A}(t)=\zeta e^{-i \omega_0 t}+\zeta^\dagger e^{i \omega_0 t}$,
makes explicit its eigenstate representation: %
$A=\sum_{\omega\in \{-\omega_0, \omega_0\}} A_{\omega}$, with
$A_{-\omega_0}=\zeta$ and $A_{\omega_0}=\zeta^{\dagger}\,$
and also the value of G=2.
In what follows, for brevity, the components $\pm \omega_0$ will be indicated simply by $\pm$ so that 
$A_{\omega_0}= \zeta^\dagger:= \zeta_+$ and 
$A_{-\omega_0}=\zeta:= \zeta_-\,.$
Under PSA we obtain the following ${\rm Shr\ddot{o}dinger}$ picture master equation:
\begin{eqnarray}
\label{eq:me-xx}
\dot{\rho}_{\rm S}(t) =
- i \left[H_{\rm S}+H_{\rm LS}^{(\Delta t)}\,,\, \rho_{\rm S}(t) \right]_{-}+\hspace{1cm} \\
\sum_{\omega {\omega '}=-, +}
\gamma_{\omega {\omega '}}^{(\Delta t)} \left\{\zeta_{\omega '}^\dagger \rho_{\rm S}(t) \zeta_\omega 
- \frac{1}{2} \left[ \zeta_\omega \zeta_{\omega '}^\dagger \,,\, \rho_{\rm S}(t)\right]_{+} \right\}\,,
\nonumber 
\end{eqnarray}
\begin{equation}
\label{eq:me-xx-LS}
H_{\rm LS}^{(\Delta t)}=\sum_{\omega {\omega '}=-, +}\eta_{\omega {\omega '}}^{(\Delta t)}  \zeta_\omega \zeta_{\omega '}^\dagger\,.
\end{equation}

According to Eqs.~(\ref{eq:S-function-def})-(\ref{eq:anti-hermitian-part-of-gammaij}) of the general formalism,
the matrices $\gamma_{\omega {\omega'}}^{(\Delta t)}=
\Gamma_{\omega {\omega'}}^{(\Delta t)}+\Gamma_{{\omega'} \omega}^{(\Delta t)\, *}$
and  $\eta_{\omega {\omega'}}^{(\Delta t)}=1/(2i)(\Gamma_{\omega {\omega'}}^{(\Delta t)}-
\Gamma_{{\omega'}  \omega}^{(\Delta t)\, *})$
are obtained from the hermitian and anti-hermitian parts of the $2\times 2$ matrix 
\begin{eqnarray}
\Gamma_{\omega {\omega'}}^{(\Delta t)}=
S_{\omega - \omega'}^{(\Delta t)}
\int_0 ^ \infty d\tau c(\tau) e^{i {\omega'} \tau} 
\,,
\label{eq:Gamma-xx}
\end{eqnarray} 
where now,  because $\omega, {\omega'} \in \{-\omega_0, \omega_0\}\,,$ we can write 
\begin{equation}
S_{\omega - \omega'}^{(\Delta t)}=
\delta_{\omega {\omega'}}+(1-\delta_{\omega {\omega'}}) {\rm sinc}(\omega_0 \Delta t)\,.
\end{equation}
As discussed at the end of the previous section the complete-positivity properties  of the master equation~(\ref{eq:me-xx}) are
directly linked to the spectrum of the matrix $\gamma^{(\Delta t)}$.
To evaluate its entries 
we pass to the continuous frequency counterpart of model,  
introducing   the density of states of the thermal bath
$D_\epsilon:=\sum_k \delta(\omega_k - \epsilon)$\,,
an the associated  spectral density
$\rho_{\epsilon}:=D_\epsilon \gamma^2_\epsilon= \sum_k \delta(\omega_k - \epsilon) \gamma_k^2$,
obtaining 
\begin{eqnarray}
\label{eq:gamma-tensor-xx}
%
 \gamma_{--}&=&\kappa_{\omega_0} n_{\omega_0}\,,  \\ \vspace{.2cm}
 \gamma_{++}&=&\kappa_{\omega_0} (1+q n_{\omega_0})\,,  
\end{eqnarray} 
which explicitly do not depends upon $\Delta t$, and 
\begin{eqnarray}  \label{DEFGAMMA-+} 
\gamma_{-+}^{(\Delta t)}&=& \gamma_{-+}^{(\Delta t)\,*}\\  \nonumber 
&=& 
\left[ \frac{( (q+1)n_{\omega_0}+ 1)  \kappa_{\omega_0}  }{2}-i \mathcal{I}\right] {\rm sinc}(\omega_0 \Delta t)  \,,
\end{eqnarray} 
where 
$\kappa_\epsilon:=2 \pi \rho_{\epsilon}$ is the system
 decay rate, and where 
%
\begin{eqnarray} 
\label{eq:principal-value-integral}
\mathcal{I}&:=&\frac{\omega_0}{\pi}
\dashint_0^\infty d\epsilon
\frac{ ((q+1) n_\epsilon + 1)\kappa_\epsilon}
{\epsilon^2 - \omega_0^2}\,, 
%
%
\end{eqnarray} 
the symbol ``$\dashint$" indicating that we are considering the principal value of the integral.
{\\
Finally we report the components of the Lamb shift matrix that will be useful in what follows: 
\begin{eqnarray}
\label{eq:lambshift-components-examples}
\eta_{--}&=&\mathcal{I}_-\,,\\
\eta_{++}&=&\mathcal{I}_+\,,\nonumber\\ 
\eta_{+-}^{(\Delta t)}&=&\left[\frac{i}{4}  (\gamma_{++}-\gamma_{--})+\frac{1}{2}(\mathcal{I}_-+\mathcal{I}_+)\right] S_{2 \omega_0}^{(\Delta t)}\,,\nonumber\\
\eta_{-+}^{(\Delta t)}&=&{\eta_{+-}^{(\Delta t)}}^*,\nonumber
\end{eqnarray}
where 
\begin{equation}
\label{eq:Ip,Im}
\mathcal{I}_{\pm}:=
\frac{1}{2 \pi}
\dashint_0^\infty d\epsilon\,
\kappa_\epsilon
\left[
\frac{1}{\epsilon\pm \omega_0} n_\epsilon
+
\frac{1}{-\epsilon\pm \omega_0} (1+q n_\epsilon)
\right]\,.
\end{equation}
}

\subsection{Critical coarse grain times as function of the bath temperature}
\label{appendix-sinc-critical}
According to the analysis of Sec.~\ref{Sec-redfield} we can ensure that 
the master Eq.~(\ref{eq:me-xx}) will describe a completely positive evolution of S  provided that 
the matrix
$\gamma^{(\Delta t)}$ is positive semidefinite, i.e. whenever its two eigenvalues 
\begin{equation}
\label{eigenvalues-of-gamma-examples}
\gamma_{\mp}^{(\Delta t)}:=\frac{1}{2}
\left( 
\gamma_{++} + \gamma_{--} \mp 
\sqrt{(\gamma_{++}- \gamma_{--})^2+ 4 |\gamma_{-+}^{(\Delta t)}|^2}
\right)\,,
\end{equation}
 are both non-negative, or equivalently
when 
\begin{eqnarray} \label{detineq} 
{\rm det}\left[ \gamma^{(\Delta t)} \right]:=\gamma_{++}\gamma_{--}-\lvert\gamma_{+-}^{(\Delta t)}\lvert^2\ge 0\,,
\end{eqnarray} 
where we used the fact that by construction $\gamma_{+}^{(\Delta t)}\geq 0$ always. 
In the SA limit, where ${\rm sinc}^2(\omega_0 \Delta t)$ approaches zero, $\gamma_{\mp}^{(\Delta t)}$ reduce
to $\gamma_{++}$ and $\gamma_{--}$ and, as anticipated in the previous section, the condition (\ref{detineq}) is always verified. 
 For small, but not zero values of ${\rm sinc}^2(\omega_0 \Delta t)$, we obtain instead first order corrections of the dissipation rates:
\begin{eqnarray} 
\gamma_-^{(\Delta t)}\simeq \gamma_{--}-\delta^{(\Delta t)}\;, \qquad \gamma_+^{(\Delta t)}\simeq \gamma_{++}+\delta^{(\Delta t)}\,,
\end{eqnarray} 
with 
\begin{eqnarray} \delta^{(\Delta t)}:=\frac{{\rm sinc}^2(\omega_0 \Delta t)}{\gamma_{++}-\gamma_{--}}  \left[\frac{(\gamma_{--}+\gamma_{++})^2}{4}+\mathcal{I}^2\right]\,.\end{eqnarray} 
Being $\gamma_{++}> \gamma_{--}$ it turns out that the presence of non-secular terms enhances the gap between the two coefficients describing dissipation.

More generally Eq.~(\ref{detineq})  leads to the following necessary and sufficient condition for the coarse graining time $\Delta t$, 
\begin{equation}\label{eq:sinc-threshold-example-ii}
| {\rm sinc}(\omega_0 \Delta t) |\leq 
\sqrt{
 \frac{4\kappa_{\omega_0}^2 n_{\omega_0} (1+q n_{\omega_0})}
 {\kappa_{\omega_0}^2[n_{\omega_0}+1+q n_{\omega_0}]^2+4\mathcal{I}^2}
}\,.
\end{equation}
Equation~(\ref{eq:sinc-threshold-example-ii}) is a necessary and sufficient condition for the positivity 
of the dynamics induced by the  markovian master equation~(\ref{eq:me-xx}).
For the $\Delta t$ values that fulfil the above inequality, 
it is instructive to put the dissipator appearing in Eq.~(\ref{eq:me-xx}) in explicit canonical GKSL form.
This is accomplished by diagonalizing the matrix $\gamma^{(\Delta t)}$ via a unitary matrix $\mathcal{U}\,:$  
$\mathcal{U}^\dagger \gamma^{(\Delta t)} \mathcal{U}={\rm diag(\gamma_-, \gamma_+)}$.
The Lindblad operators are then given by the rotated ladder operators 
\begin{eqnarray}
f_-:= \mathcal{U}_{--} \zeta + \mathcal{U}_{+-} \zeta^\dagger, \quad 
f_+:= \mathcal{U}_{-+} \zeta + \mathcal{U}_{++} \zeta^\dagger,
\end{eqnarray} 
and the dissipator becomes
\begin{eqnarray}
\mathcal{D}^{(\Delta t)} [\cdots]&=&\gamma_- \left(f_-^\dagger  \cdots f_- -\frac{1}{2} \Big[ f_- f_-^\dagger,  \cdots \Big]_+\right)\nonumber\\
&+& \gamma_+ \left(f_+^\dagger \cdots f_+-\frac{1}{2}  \Big[f_+ f_+^\dagger,  \cdots \Big]_+\right)\,, \label{DISSI} 
\end{eqnarray} 
where for easy of notation we dropped any reference to the functional dependence upon $\Delta t$ of the terms that appears on the right-hand-side term. 
It is worth noticing that in general for finite  values of $\Delta t$, one has that
 $f_+$ is not the adjoint counterpart of $f_-$, i.e. $f_+ \neq f_-^\dagger$, 
  the identity instead holding  in the full SA limit where $\mathcal{U}=\mathbb{1}\,.$

\begin{figure}
\vspace{.5cm}
\begin{overpic}[width=0.49\linewidth]{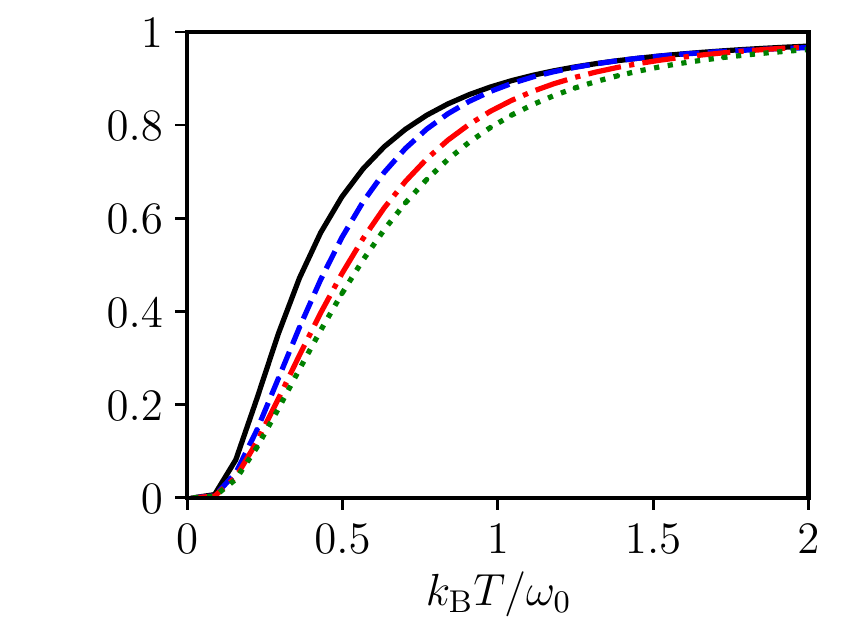}\put(5,75){(a)}\put(30,3){}\end{overpic}
\begin{overpic}[width=0.49\linewidth]{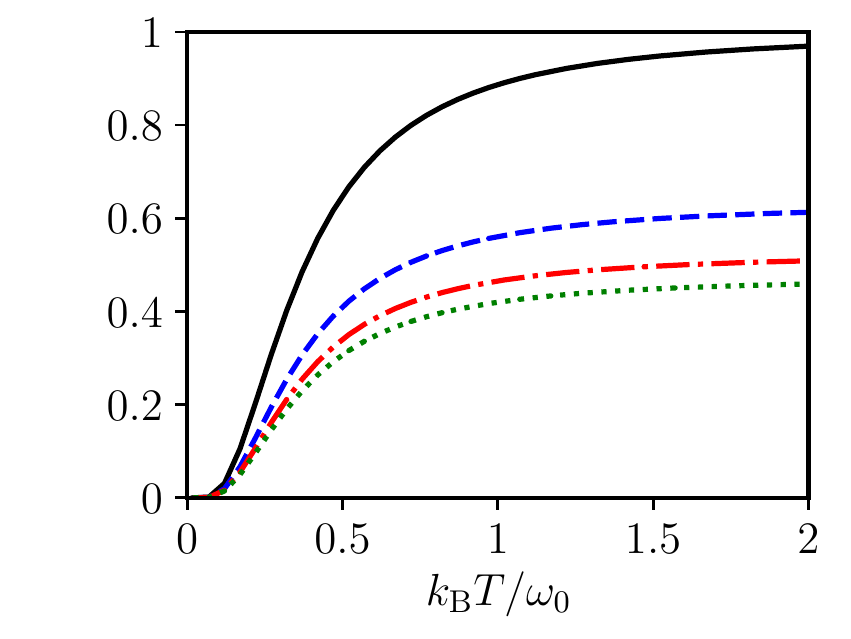}\put(5,75){(b)}\put(30,3){}\end{overpic}
\caption{(Color online) 
Plot of the r.h.s. term of
Eq.~(\ref{eq:sinc-threshold-example-ii})
 as function of temperature (the last in units $\omega_0/k_{\rm B}$). 
 Complete positivity gets lost for the PSA master equation~(\ref{eq:me-xx}) for coarse graining times $\Delta t$ 
associated with values of $| {\rm sinc}(\omega_0 \Delta t)|$ that lie above this curve. 
In Panel (a) we plot the results for a bosonic bath by choosing different values of $\omega_c$: 
$\omega_c=10 \omega_0$  (blue dashed line);
$\omega_c=20 \omega_0$  (red dot-dashed line);
$\omega_c=30 \omega_0$  (green dotted line). 
The black full line reports instead the function
$\frac{2\sqrt{ n_{\omega_0} (1+q n_{\omega_0})}}
 {n_{\omega_0}+1+q n_{\omega_0}}$
 which provide a simple upper bound for the r.h.s.
 Eq.~(\ref{eq:sinc-threshold-example-ii})  obtained by
 by setting ${\cal I}=0$ in the latter. 
Panel (b): same as in Panel (a) but for a fermionic bath. 
The adimensional constant $\kappa_0$ of Eq.~(\ref{eq:decay-rate-exp-cutoff}) is irrelevant in this analysis.
\label{fig:delta-toy-models}}
\end{figure}

\begin{figure}
\begin{overpic}[width=0.49\linewidth]{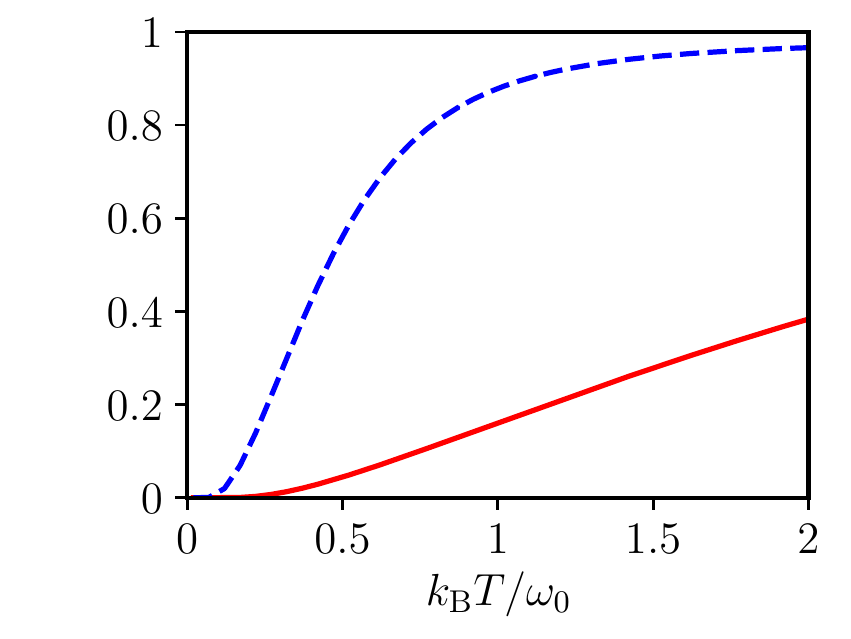}\put(5,75){(a)}\put(30,3){}\end{overpic}
\begin{overpic}[width=0.49\linewidth]{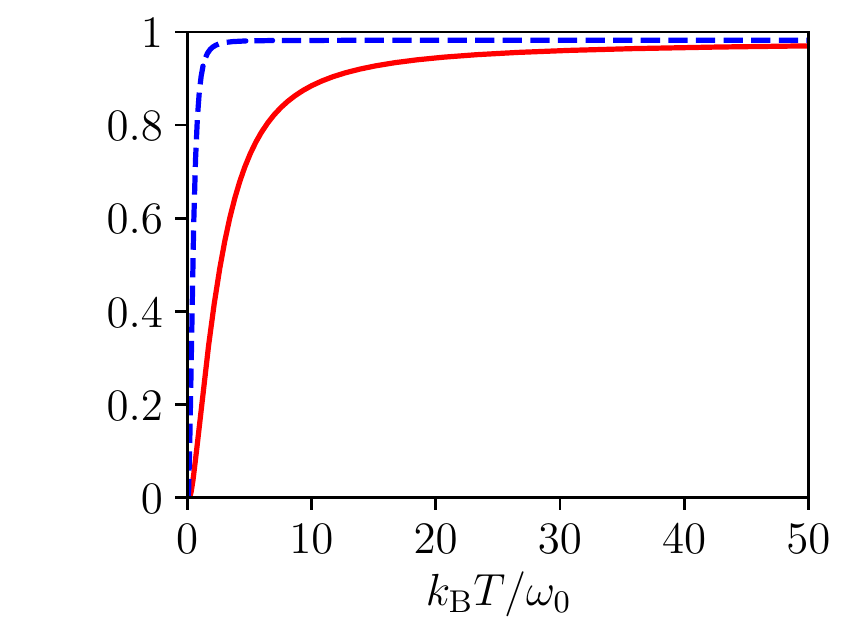}\put(5,75){(b)}\put(30,3){}\end{overpic}
\begin{overpic}[width=0.49\linewidth]{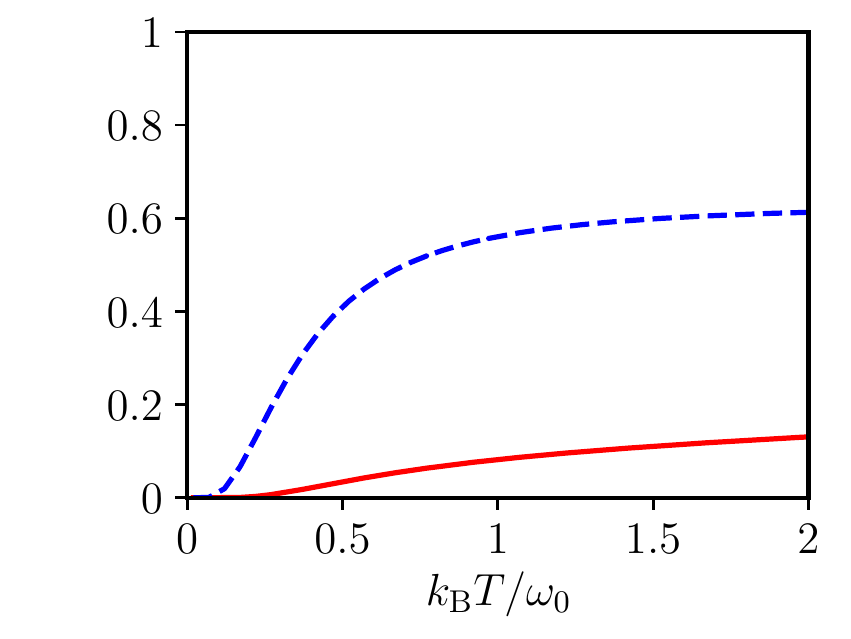}\put(5,75){(c)}\put(30,3){}\end{overpic}
\begin{overpic}[width=0.49\linewidth]{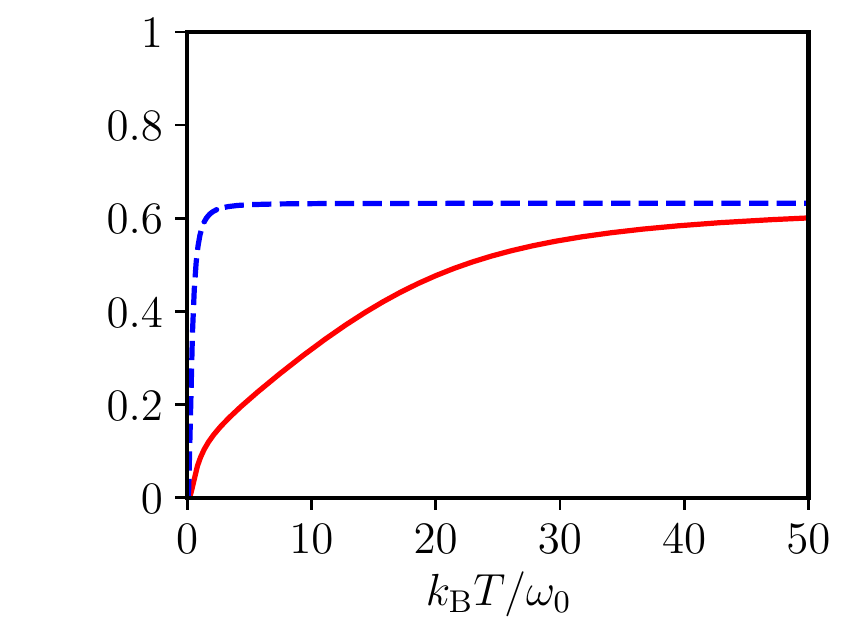}\put(5,75){(d)}\put(30,3){}\end{overpic}
\caption{(Color online) 
Panel (a): comparison between the
r.h.s. term of
Eq.~(\ref{eq:sinc-threshold-example-ii}) 
 as function of temperature  (blue dashed line) 
and the r.h.s. term of
 Eq.~(\ref{eq:bound-example1})  (red full line) for a bosonic bath.
Panel (b): same as Panel (a) but with a different range of temperature.
Panels (c) and (d): same as in Panels (a) and (b) but for a fermionic bath. 
The cutoff energy has been fixed at $\omega_c=10 \omega_0$ for all the curves.
\label{fig:delta-bound}}
\end{figure}
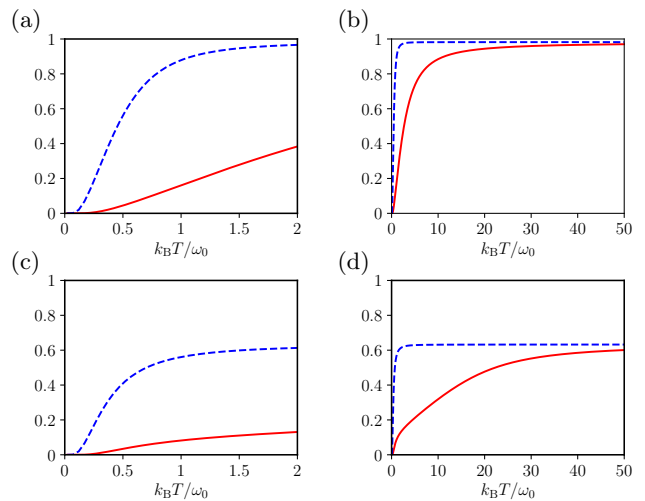

A plot of the upper bound on the r.h.s. of the above expressions is reported in Fig.~\ref{fig:delta-toy-models} as function of $k_{\rm B} T/\omega_0$, both for bosonic [Panel (a)] and for fermionic baths [Panel (b)] for a decay rate of the form 
\begin{equation}
\label{eq:decay-rate-exp-cutoff}
\kappa_\epsilon= \kappa_0 \epsilon \exp(-\epsilon/\omega_c)\,. 
\end{equation}
that behaves ohmically for small energies --- $\kappa_\epsilon\sim \epsilon$ for $\epsilon\ll \omega_c$ ---  and decays exponentially with the cutoff energy $\omega_c\,.$
From Fig.~\ref{fig:delta-toy-models} we deduce that
at low temperature the full SA ($\Delta t\rightarrow \infty$) is necessary for ensuring positivity; 
this is a general behaviour that does not depend upon the special form of the decay rate we choose for the plot. Indeed
for $\beta\rightarrow \infty$ the right-hand side of Eq.~(\ref{eq:sinc-threshold-example-ii}) always nullifies forcing us to take $\Delta t\rightarrow \infty$ in order to 
satisfy the inequality; for non-zero temperature values instead, finite values of $\Delta t$ are admitted such that the associated PSA master equation is well behaved.
Furthermore we note that the only difference between the bosonic (Fig.~\ref{fig:delta-toy-models}a) and the fermionic (Fig.~\ref{fig:delta-toy-models}b) cases comes just from the principal value integral $\mathcal{I}\,$. The last turns out to be independent of temperature only for fermions, see Eq.~(\ref{eq:principal-value-integral}).
Finally in Fig.~\ref{fig:delta-bound} we compare the actual threshold value of Eq.~(\ref{eq:sinc-threshold-example-ii}) with the value provided 
by the estimation of Eq.~(\ref{eq:sinc-upper-bound-norm-and-eigen}) associated with the sufficient positivity condition we derive in 
Sec.~\ref{SEC.iv} under general assumption on the system dynamics, which for examples we study here 
 assumes the form 
\begin{eqnarray}
\label{eq:bound-example1}
 |{\rm sinc}(\omega_0 \Delta t) |\leq 
\tfrac
{
2 \kappa_{\omega_0} n_{\omega_0}
}
{
\sqrt{\left[ \kappa_{\omega_0} n_{\omega_0}\right]^2+4\mathcal{I}_-^2}+
\sqrt{\left[ \kappa_{\omega_0} (1 + q n_{\omega_0})\right]^2+4\mathcal{I}_+^2}
}\,,
\nonumber
\\
\end{eqnarray}
where $\mathcal{I}_\pm$ are the same as in Eq.(\ref{eq:Ip,Im}) 
[notice that $\mathcal{I}=\mathcal{I}_{-}-\mathcal{I}_{+}\,,$ see Eq.~(\ref{eq:principal-value-integral})].
We infer that Eq.~(\ref{eq:bound-example1}) underestimates the threshold value at low temperatures and gives better results at high temperatures.

\subsection{Commutativity} \label{sec:comm} 
As anticipated in the end of Sec.~\ref{Sec-redfield} 
the main structural consequence of the implementation of the PSA is the breaking of the commutation rules either at the level of the operators than at the level of the super-operators entering in Eq.~(\ref{eq:me-xx}).
We distinguish now the cases of qubit and QHO and in particular we concentrate on three main aspects:
i) the value of the commutator between the Lamb shift and the free Hamiltonian of the system, 
ii) the change of the commutation rules at the level of super-operators,  
iii) the effects of the non-secular terms in the dynamics.  
About item ii) we remark that both for qubit and for QHO we are in the hypotheses of Eq.~(\ref{labshift-eigenoperator}) for which Eq.~(\ref{H-D-commute}) is satisfied.

\subsubsection{Qubit}
In the case of qubit $(s=-1)$ the ladder operator $\zeta$ is the operator $\sigma_-=\ket{0}\bra{1}\,,$ with $\ket{0}$ and $\ket{1}$ being the eigenvectors of the Pauli matrix $\sigma_z$ corresponding to the eigenvalues $-1$ and $1$, respectively. Because $\sigma_\pm^2=0\,,$  the Hamiltonian of the system is modified just by a change of the two level spacing of the system and is independent of $\Delta t\,,$ i.e. 
\begin{eqnarray} 
H_{\rm S}^{(\Delta t)} &=& \bar{\omega} \; \sigma_+ \sigma_- \,, \\ 
 \bar{\omega}&:=&\omega_0 + \eta_{++}-\eta_{--}\,,  \label{OMEGABAR} 
\end{eqnarray} 
with $\eta_{--}$ and $\eta_{++}$ as in Eq.~(\ref{eq:lambshift-components-examples}). 
Accordingly for all allowed PSA values $\Delta t$, this grants commutation between the free and LS contributions of the full Hamiltonian, i.e. 
\begin{equation}
[H_{\rm LS}^{(\Delta t)}, H_{\rm S}]_-=0\,.
\end{equation}
On the contrary the dissipator (\ref{eq:me-xx})  acquires non-secular terms which depend on $\Delta t\,:$
\begin{eqnarray}
\mathcal{D}^{(\Delta t)}[\rho_{\rm S}(t)]=
\gamma_{--} \left(\sigma_+ \rho_{\rm S}(t) \sigma_- 
- \frac{1}{2} \left\{\sigma_- \sigma_+  \,,\, \rho_{\rm S}(t)\right\} \right)\nonumber \\
+ 
\gamma_{++} \left(\sigma_- \rho_{\rm S}(t) \sigma_+ 
- \frac{1}{2} \left\{\sigma_+ \sigma_-  \,,\, \rho_{\rm S}(t)\right\} \right)\nonumber\\
+
\gamma_{- +} ^{(\Delta t)} \sigma_- \rho_{\rm S}(t) \sigma_-
+
\gamma_{+ -} ^{(\Delta t)} \sigma_+ \rho_{\rm S}(t) \sigma_+ 
\,, \nonumber 
\end{eqnarray} 
with $\gamma_{+ -} ^{(\Delta t)}$ as in Eq.~(\ref{DEFGAMMA-+}).
Hence at the level of super-operators, it is interesting
to observe that the Hamiltonian and the dissipating parts of the generator of the dynamical semi-group don't commute under PSA:
\begin{equation}
\label{qubit-commutator-H-D-II}
\left[\mathcal{D}^{(\Delta t)} ,  \mathcal{H}_{\rm S}^{(\Delta t)} \right]_-=
-2i \bar{\omega}
\left(
\gamma_{-+}^{(\Delta t)}
\sigma_- \rho \sigma_-
-{\rm h.c}
\right)
\,.
\end{equation}
Notice that consistently the commutator in Eq.~(\ref{qubit-commutator-H-D-II}) nullifies in the limit 
$\Delta t \rightarrow \infty$, i.e. within the SA. Analogous considerations hold about the term  $\left[\, \mathcal{D}^{(\Delta t)}, \mathcal{H}_{\rm S}\right]_-$ once replaced $\bar{\omega}$ with the bare frequency $\omega_0\,.$

\begin{figure}
\vspace{.5cm}
\begin{overpic}[width=0.49\linewidth]{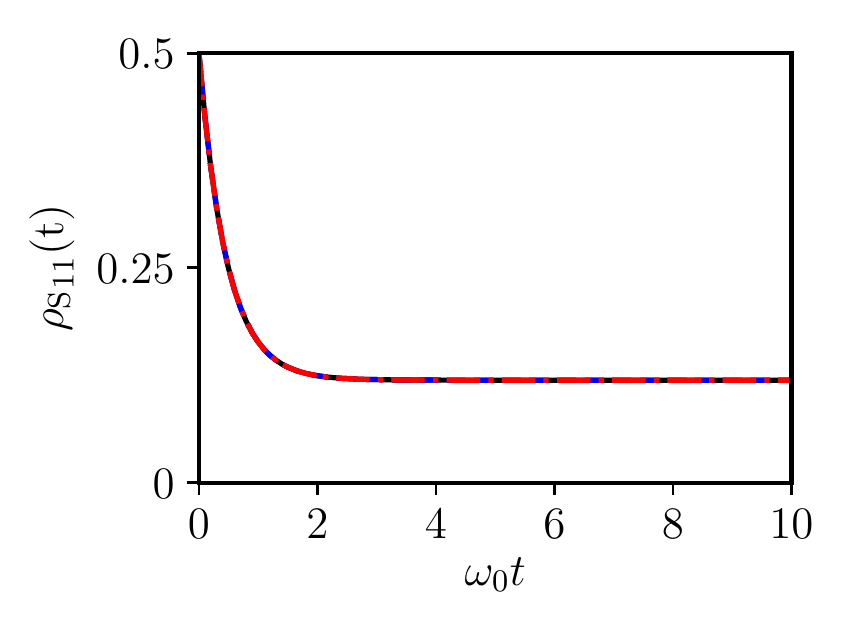}\put(5,75){(a)}\put(30,3){}\end{overpic}
\begin{overpic}[width=0.49\linewidth]{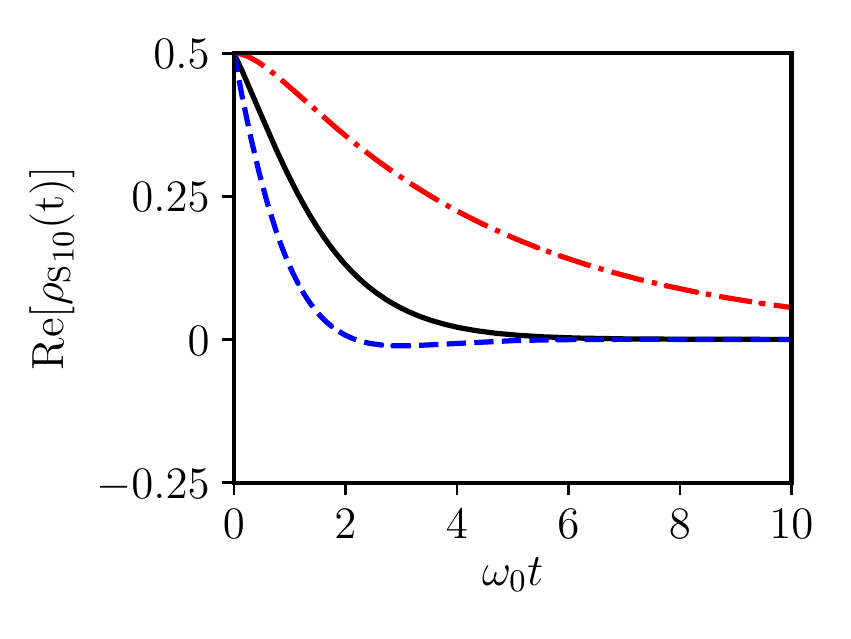}\put(5,75){(b)}\put(30,3){}\end{overpic}
\begin{overpic}[width=0.49\linewidth]{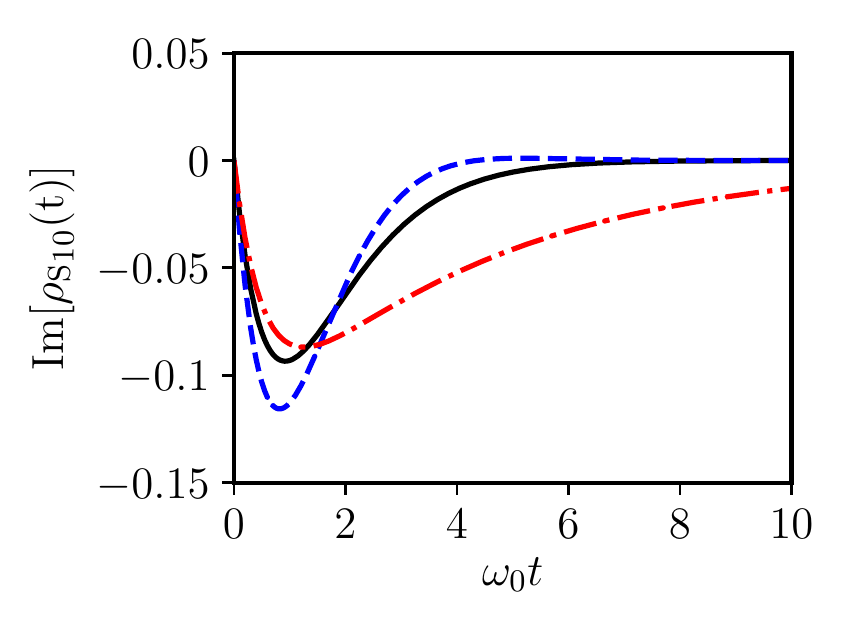}\put(5,75){(c)}\put(30,3){}\end{overpic}
\begin{overpic}[width=0.49\linewidth]{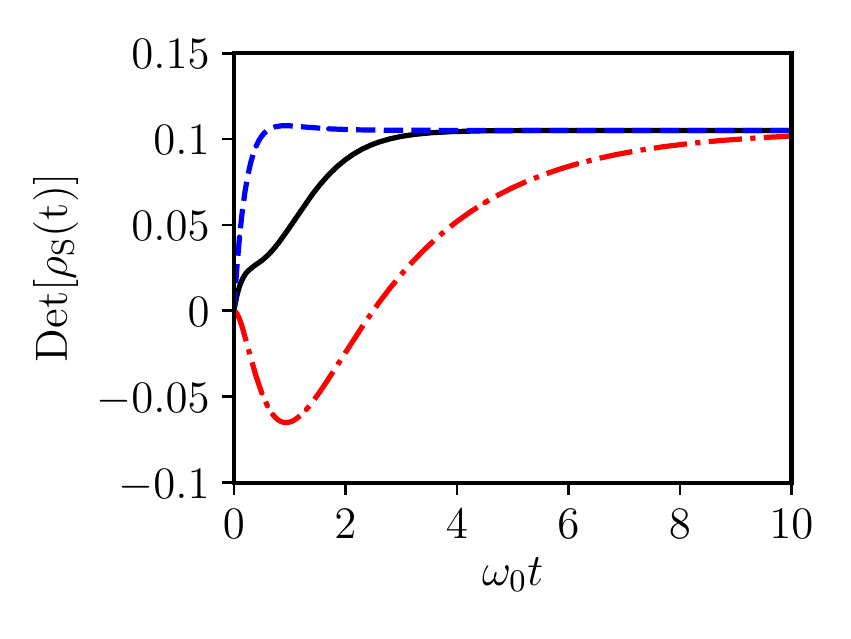}\put(5,75){(d)}\put(30,3){}\end{overpic}
\caption{(Color online) 
Qubit interacting with a bosonic bath ($q=1$):
We plot ${\rho_{\rm S}}_{11}(t)$ in Panel (a), ${\rm Re}{[\rho_{\rm S}}_{10}(t)]$ in Panel (b), ${\rm Im}{[\rho_{\rm S}}_{10}(t)]$ in Panel (c) and ${\rm Det}[{\rho_{\rm S}}(t)]$ in Panel (d) as function of time in units $1/\omega_0$.
We choose the following values of the master equation parameters:
$k_{\rm B} T= 0.5 \omega_0$, $\kappa_0=2$,  $\omega_c=5 \omega_0$, see Eq.~(\ref{eq:decay-rate-exp-cutoff}).
In all the Panels the black full lines correspond to the results obtained using PSA at positivity threshold ($ {S_{2 \omega_0}^{(\Delta t)}} \approx 0.628.$), the dash-dotted red lines using Redfield ($ {S_{2 \omega_0}^{(\Delta t)}} =1$) and the blue dashed line using SA
($ {S_{2 \omega_0}^{(\Delta t)}} =0$).
%
\label{fig:positivity-evolution}}
\end{figure}
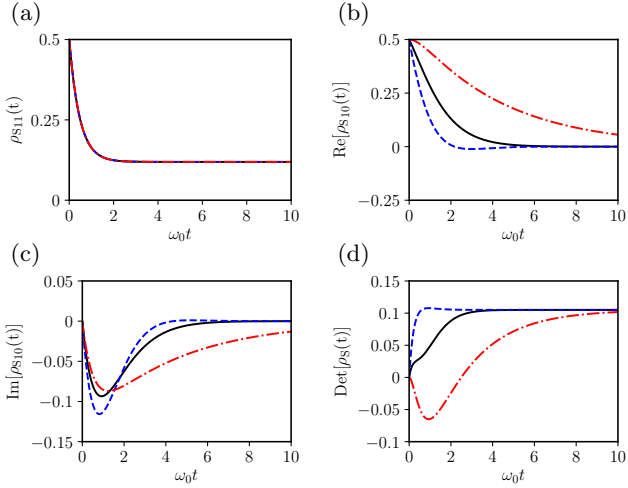

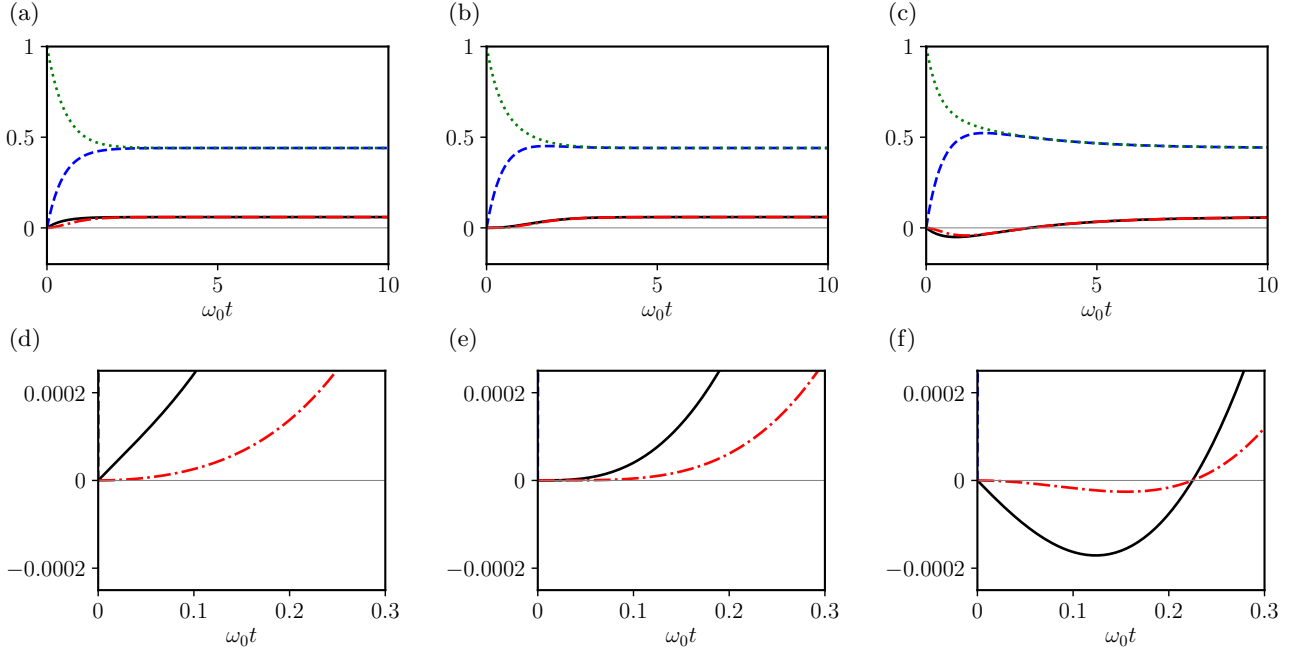
\begin{figure*}
\vspace{.5cm}
\begin{overpic}[width=0.32\linewidth]{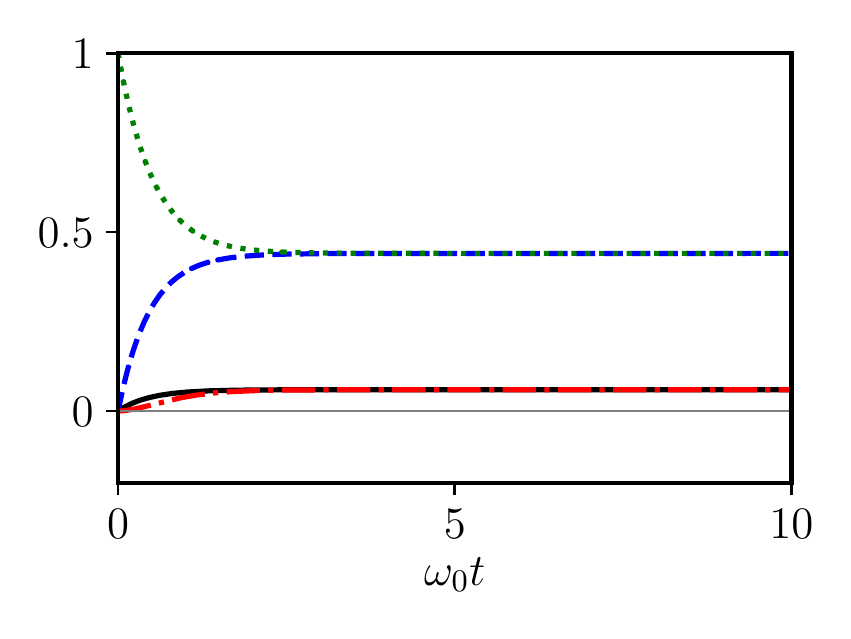}\put(5,75){(a)}\put(30,3){}\end{overpic}
\begin{overpic}[width=0.32\linewidth]{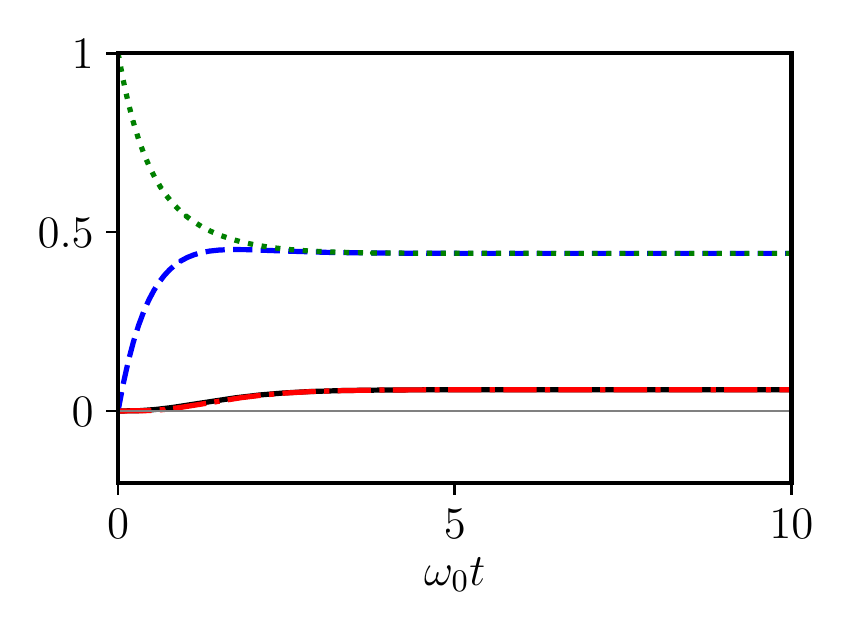}\put(5,75){(b)}\put(30,3){}\end{overpic}
\begin{overpic}[width=0.32\linewidth]{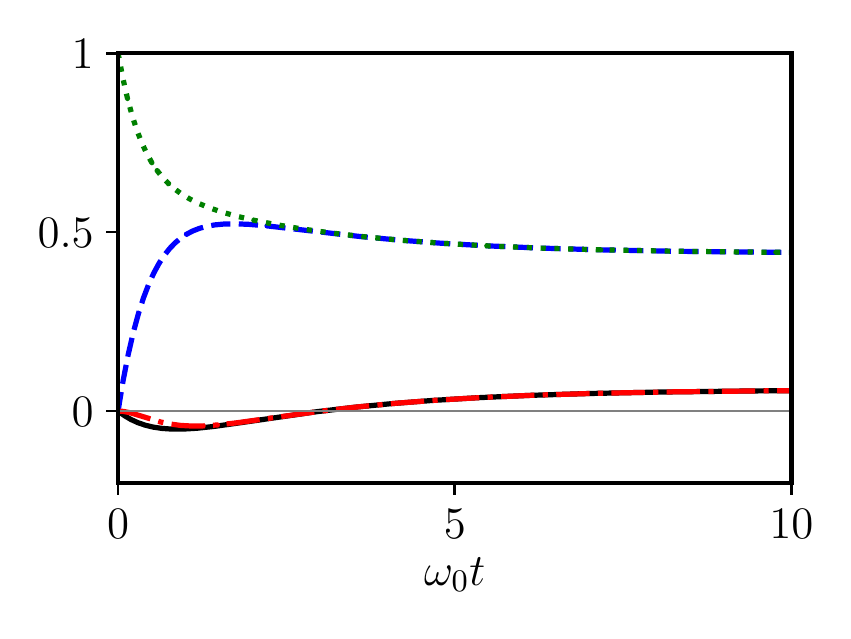}\put(5,75){(c)}\put(30,3){}\end{overpic}
\begin{overpic}[width=0.32\linewidth]{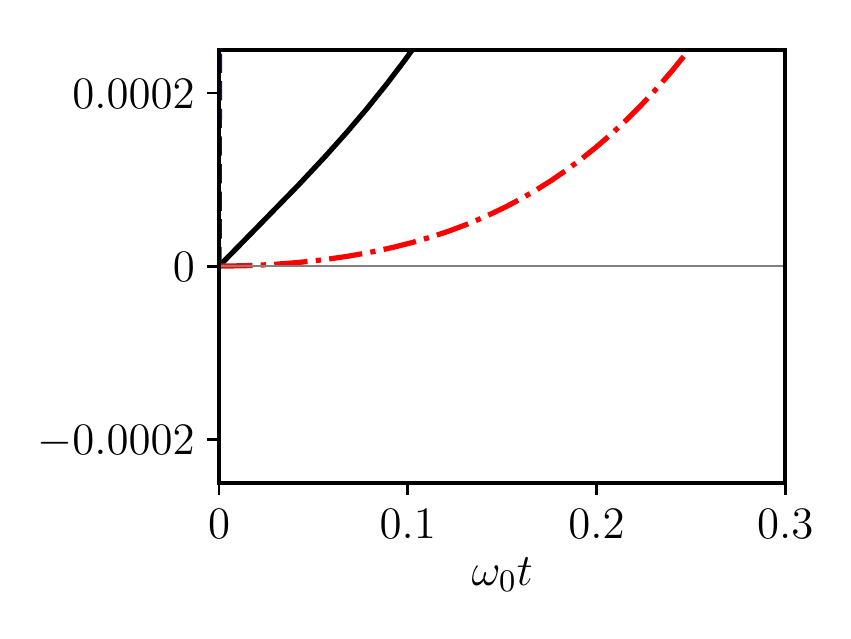}\put(5,75){(d)}\put(30,3){}\end{overpic}
\begin{overpic}[width=0.32\linewidth]{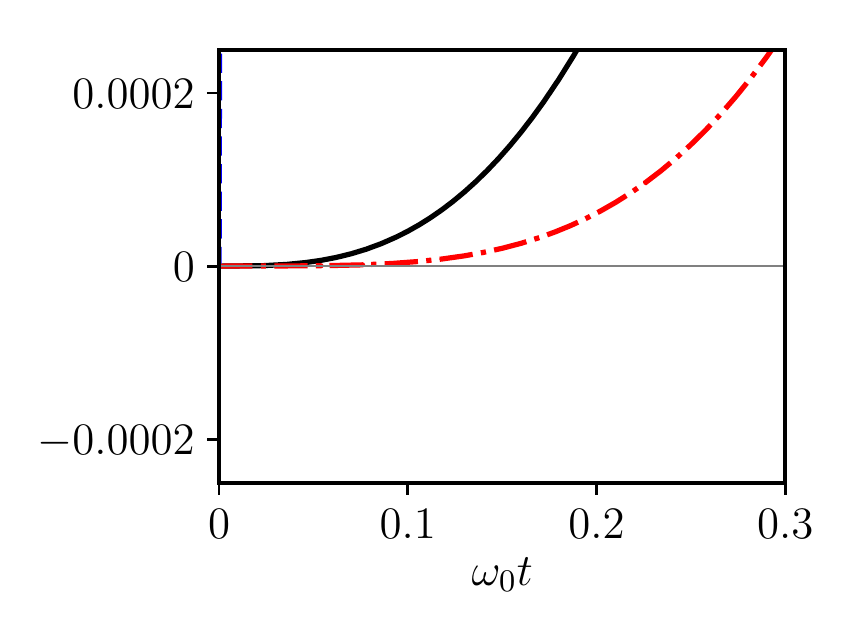}\put(5,75){(e)}\put(30,3){}\end{overpic}
\begin{overpic}[width=0.32\linewidth]{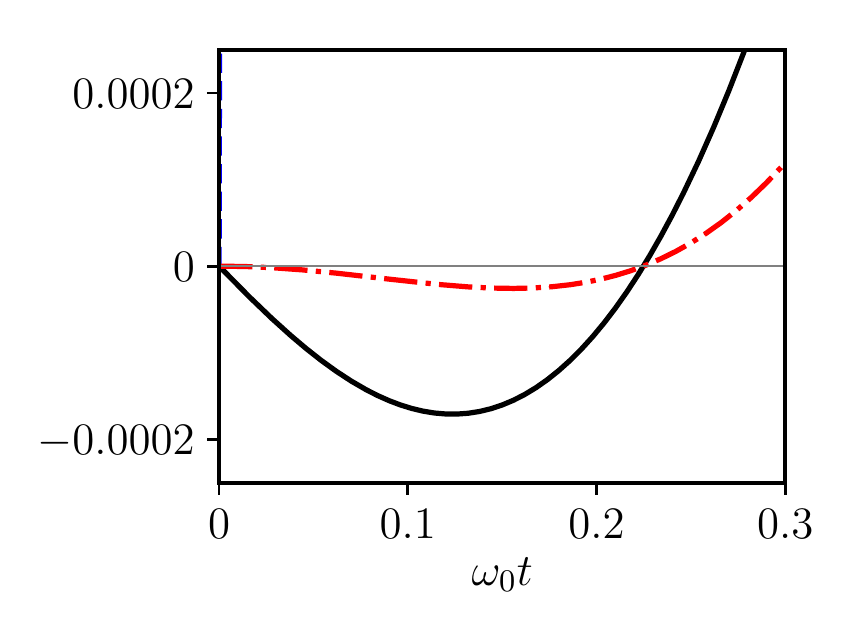}\put(5,75){(f)}\put(30,3){}\end{overpic}
\caption{(Color online) 
Qubit interacting with a bosonic bath ($q=1$).
We plot the four eigenvalues of $\rho_{\rm CJ}(t)$ as function of time in units $1/\omega_0$ for different values of ${S_{2 \omega_0}^{(\Delta t)}}$.
Panel (a): ${S_{2 \omega_0}^{(\Delta t)}}=0$ (SA);
Panel (b): ${S_{2 \omega_0}^{(\Delta t)}}=0.628$ (PSA at positivity threshold ${S_{2 \omega_0}^{(\Delta t)}}={S_{2 \omega_0}^{(\Delta t)}}_{\rm th}$);
Panel (c): ${S_{2 \omega_0}^{(\Delta t)}}=1 $ (Redfield).
Then we move across the threshold value to appreciate the crossover between CP and non-CP evolution.
Panel (d): ${S_{2 \omega_0}^{(\Delta t)}}=0.621$
$({S_{2 \omega_0}^{(\Delta t)}}$ slightly below $ {S_{2 \omega_0}^{(\Delta t)}}_{\rm th})\,;$ 
Panel (e): ${S_{2 \omega_0}^{(\Delta t)}}=0.628$
$({S_{2 \omega_0}^{(\Delta t)}}= {S_{2 \omega_0}^{(\Delta t)}}_{\rm th})\,;$ 
Panel (f): ${S_{2 \omega_0}^{(\Delta t)}}=0.634$
$({S_{2 \omega_0}^{(\Delta t)}}$ slightly above $ {S_{2 \omega_0}^{(\Delta t)}}_{\rm th})\,.$
Notice the different axes scales with respect to the first three Panels.
We choose the following values of the master equation parameters:
$k_{\rm B} T= 0.5 \omega_0$, $\kappa_0=2$,  $\omega_c=5 \omega_0$.
%
%
\label{fig:choi}}
\end{figure*}

We show now an example of non-positive semi-definite evolution by considering as initial state the pure vector $|\psi(0)\rangle_S :=( |0\rangle+|1\rangle)/\sqrt{2}$.
%
%
Its dynamics is described by Eq.(\ref{eq:me-xx}), which in a more explicit form reads as
\begin{eqnarray}
\frac{d}{dt} {\rho_{\rm S}}_{00} &=& \gamma_{++} \left(1 - {\rho_{\rm S}}_{00}\right) - \gamma_{--} {\rho_{\rm S}}_{00}, \\
\frac{d}{dt}{\rho_{\rm S}}_{10}&=&-i \bar{\omega} {\rho_{\rm S}}_{10} 
-\frac{1}{2}\left(\gamma_{++}+\gamma_{--}\right)
{\rho_{\rm S}}_{10}
+ \gamma_{+-}^{\left(\Delta t\right)} {\rho_{\rm S}}_{10}^*
\,.
\nonumber
\end{eqnarray}
We obtain the following analytic expressions of the components of the density matrix ${\rho_{\rm S}}(t)$:
\begin{widetext}
\begin{eqnarray}
{\rm Re}[ {\rho_{\rm S}}_{10}](t)&=&{\rm Re}[ {\rho_{\rm S}}_{01}](t) =\frac{1}{2} e^{-\frac{1}{2}s t}
 \left(\tfrac{
{\rm Re}\left[\gamma_{-+}^{(\Delta t)}\right] 
\sin \left(  {\bar{\omega}_{\Delta t}}\,t\right)}
{ {\bar{\omega}_{\Delta t}}}+
\cos \left( {\bar{\omega}_{\Delta t}}\,t\right)\right)
\,,
\\
{\rm Im }[{\rho_{\rm S}}_{10}](t)&=&-{\rm Im }[{\rho_{\rm S}}_{01}](t) = -\tfrac{1}{2}e^{-\frac{1}{2}s t} \frac{
\left({\rm Im}\left[\gamma_{-+}^{(\Delta t)}\right]+{\bar{\omega}} \right)  
\sin \left(\bar{\omega}_{\Delta t}\,t\right)}
{ {\bar{\omega}_{\Delta t}}}
\,,
\\
{\rho_{\rm S}}_{00}(t)&=& 1-{\rho_{\rm S}}_{11}(t) = \frac{-d e^{-st}+2 \gamma_{++}}{2 s}
\,,
\end{eqnarray}
\end{widetext}
with
\begin{eqnarray}
&s:=\gamma_{++}+\gamma_{--}\,, \qquad  \nonumber 
d:=\gamma_{++}-\gamma_{--}\,,&\\
&{\bar{\omega}_{\Delta t}}:=\sqrt{\bar{\omega}^2-|{\gamma_{-+}^{(\Delta t)}}|^2}\,.& \label{defnewnew} 
\end{eqnarray}
We plot the results in Fig.~\ref{fig:positivity-evolution} for different values of $S_{2 \omega_0}^{(\Delta t)}$ corresponding to the SA (i.e. $S_{2 \omega_0}^{(\Delta t)}=0$);  to the PSA at positivity threshold (bound of Eq.~(\ref{eq:sinc-threshold-example-ii})); and to the Redfield regime (i.e. $S_{2 \omega_0}^{(\Delta t)}=1$).  
Panels (a), (b) and (c) of the figure  show that the PSA --- when compared to the SA --- implies corrections on the off-diagonal terms of $\rho_{\rm S}(t)$ only, while leaving unchanged the diagonal ones and the steady state of the system, i.e.  
%
\begin{equation}
\label{eq:gibbs-qubit}
\rho_{\rm S}(\infty)={\rho_\beta}_{\rm S}:=
\begin{pmatrix}
n_f(\omega_0) & 0\\
0& 1-n_f(\omega_0)\\
\end{pmatrix}\,,
\end{equation}
where $n_f(\omega_0)$ is the Fermi-Dirac occupation number,
$n_f(\omega_0):=\frac{1}{e^{\beta \omega_0}+1}\,.$
As evident from the plots PSA somehow interpolates between SA and the R-B behaviours, retaining part of the
fast oscillations of the latter which are instead washed away by the former. 
In Panel (d) of Fig.~\ref{fig:positivity-evolution}   
it is instead plotted the determinant of $\rho_{\rm S}(t)$. For short time scales Redfield implies non-positive evolution being $\rm Det[\rho_{\rm S}(t)]<0$, whilst positivity is maintained under SA and PSA.  
%
%
%
%
%
%
%
%
%
%
%
%
To study the complete positivity of the system evolution we use the channel-state duality provided by the Choi-Jamio\l{}kowski  (CJ) isomorphism~\cite{HOLEVOBOOK,CHOI,JAM}.
Given a quantum channel $\Phi_{\rm S}(t)$ which maps $\rho_{\rm S}(0)\rightarrow \rho_{\rm S}(t)$, considering the situation in which S is maximally entangled with an ancilla A having the same dimensionality as S, the CJ-state $\rho_{\rm CJ}(t)$ is obtained by applying the channel locally on S.
In this case it reads as 
\begin{eqnarray}
\rho_{\rm CJ}(t):=(\Phi_{\rm S}(t) \otimes id_{\rm A}) (\ket{\psi}\bra{\psi}_{\rm SA})\,, 
\end{eqnarray}
where the initial state $\ket{\psi}_{\rm SA}$ is the maximally entangled state
\begin{eqnarray}
\ket{\psi}_{\rm SA}:=\frac{1}{\sqrt{2}}(\ket{00}_{\rm SA}+\ket{11}_{\rm SA})\,.
\end{eqnarray}
Hence $\rho_{\rm CJ}(t)$ is a $4 \times 4$-matrix having the following block form:
\begin{eqnarray}
\rho_{\rm CJ}(t)=
\frac{1}{2}
\begin{pmatrix}
\rho_{\rm S}(t;\ket{1}\bra{1})& \rho_{\rm S}(t;\ket{1}\bra{0})\\ 
\rho_{\rm S}(t;\ket{0}\bra{1}) & \rho_{\rm S}(t;\ket{0}\bra{0})
\end{pmatrix}\,,
\end{eqnarray} 
with $\rho_{\rm S}(t;\ket{i}\bra{j})$ being the $2 \times 2$-matrix obtained by the solution of the ME Eq.~(\ref{eq:me-xx}) under the initial condition $\rho_{\rm S}(0)=\ket{i}\bra{j}\,.$
The occurrence of negative eigenvalues of $\rho_{\rm CJ}(t)$ encodes the non-complete positivity (non-CP) of the map $\Phi_{\rm S}(t)$.
In Fig.~\ref{fig:choi} we plot the four eigenvalues of $\rho_{\rm CJ}(t)$ as function of time for different values of ${S_{2 \omega_0}^{(\Delta t)}}\,.$ Non-CP manifests at short time scales as soon as the threshold value of Eq.~(\ref{eq:sinc-threshold-example-ii}) is overcome. 
This can be understood by looking at the analytic expression of the eigenvalue ${\lambda^{(\Delta t)}}(t)$ corresponding to the black full lines in 
Fig.~\ref{fig:choi}:
\begin{widetext}
{

\begin{eqnarray}
\label{eq:choi-eigenvalue}
{\lambda^{(\Delta t)}}(t)= 
\frac{1}{4} 
\left[1-e^{-st}-
\frac{
\sqrt{2}}{s {\bar{\omega}_{\Delta t}} }
e^{-st/2} 
\sqrt{
d^2 {\bar{\omega}_{\Delta t}}^2 \left(\cosh (s t)-1\right)-
|{\gamma_{-+}^{(\Delta t)}}|^2
s^2 \left(\cos (2  {\bar{\omega}_{\Delta t}}  t)-1\right)
}
\right]\,,
\end{eqnarray}
}
\end{widetext}
where the parameters $d$, $s$ and  ${\bar{\omega}_{\Delta t}}$ are again defined as in Eq.~(\ref{defnewnew}). 
Being
\begin{eqnarray}
\label{eq:choi-derivative}
{\lambda^{(\Delta t)}}(0)&=&0\,,\\
{\dot{\lambda}^{(\Delta t)}}(0)&=&
\frac{1}{4}\left(s-\sqrt{d^2+4 {|\gamma_{-+}^{(\Delta t)}|}^2}\right)\,,
\end{eqnarray}
we obtain that
the first derivative ${\dot{\lambda}^{(\Delta t)}}(0)\gtrless 0$ when 
$|\gamma_{-+}^{(\Delta t)}|^2\lessgtr \gamma_{--}\gamma_{++}$
and, consequently, at short time scales it happens that  
$\lambda_1(t)\gtrless 0$.
About the steady state, for any value of ${S_{2 \omega_0}^{(\Delta t)}}\,,$ 
$\rho_{\rm CJ}(\infty)= {\rho_\beta}_{\rm S}\otimes \frac{1}{2} \mathbb{1}_{\rm A} $ with
${\rho_\beta}_{\rm S}$ being again the Gibbsian state of Eq.~(\ref{eq:gibbs-qubit}).
\subsubsection{Quantum harmonic oscillator}
\begin{figure}
\vspace{.5cm}
\begin{overpic}[width=0.49\linewidth]{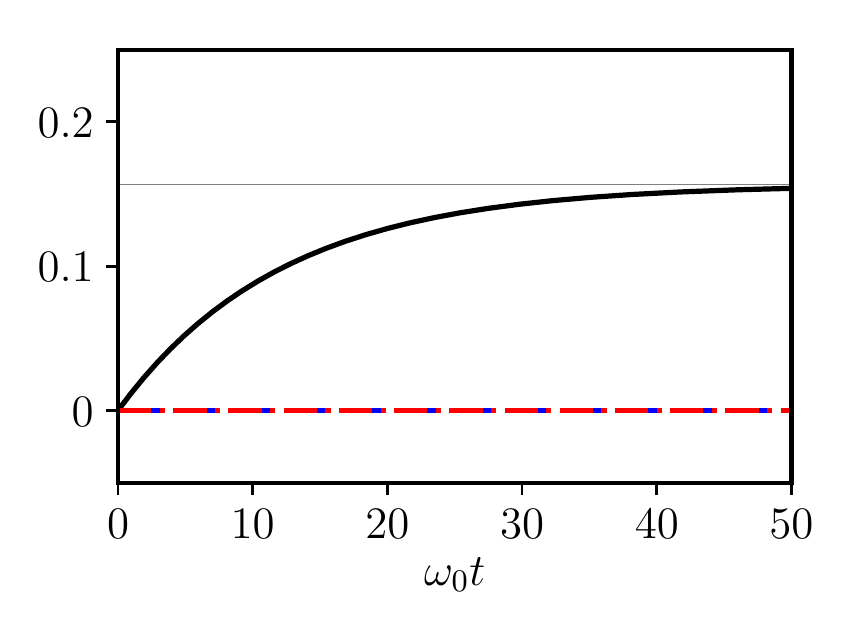}\put(5,75){(a)}\put(30,3){}\end{overpic}
\begin{overpic}[width=0.49\linewidth]{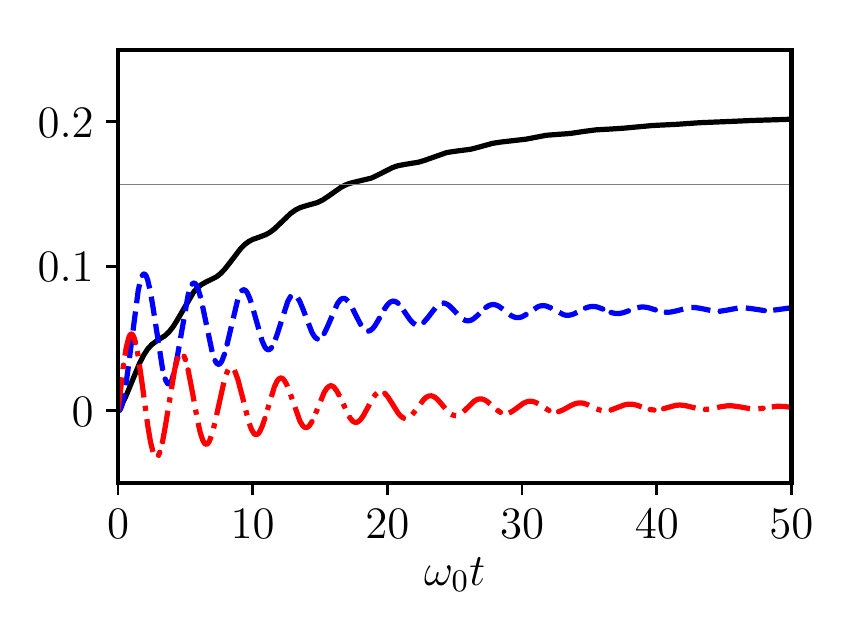}\put(5,75){(b)}\put(30,3){}\end{overpic}
\begin{overpic}[width=0.49\linewidth]{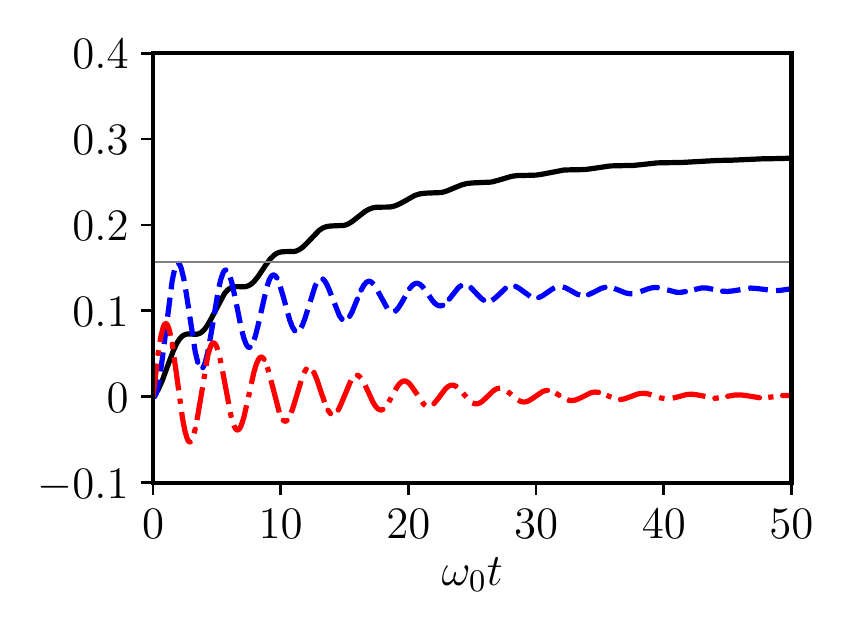}\put(5,75){(c)}\put(30,3){}\end{overpic}
\begin{overpic}[width=0.49\linewidth]{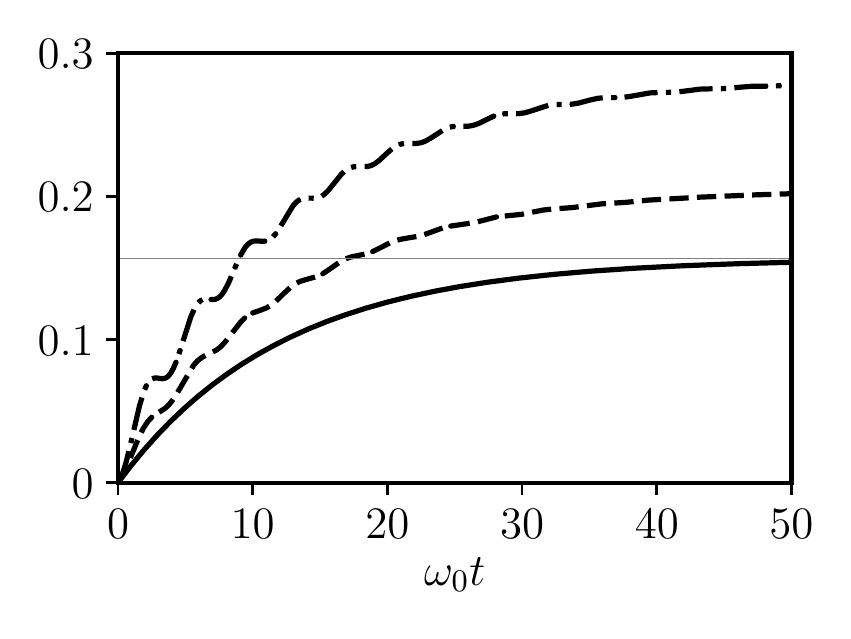}\put(5,75){(d)}\put(30,3){}\end{overpic}
\caption{(Color online) 
Quantum harmonic oscillator interacting with a bosonic bath ($q=1$).
We consider as initial state the ground state  $\rho_{\rm S}(0)=\ket{0}\bra{0}$ of $H_{\rm S}\,.$
Panel (a): we plot $\langle a^\dag a \rangle$ (black full line), ${\rm Re} \langle a^2 \rangle$ (blue dashed line), ${\rm Im} \langle a^2 \rangle$ (red dot-dashed line) as function of time in units $1/\omega_0$ in the SA ($S_{2 \omega_0}^{(\Delta t)}=0$).
Panel (b): same as Panel (a) but in the PSA at positivity threshold ($S_{2 \omega_0}^{(\Delta t)}={S_{2 \omega_0}^{(\Delta t)}}_{\rm th}=0.628$).
Panel (c): same as Panel (a) and (b) but using Redfield ($S_{2 \omega_0}^{(\Delta t)}=1$).
In Panel (d) we compare the occupation number $\langle a^\dag a\rangle $ in the three cases:
SA (black continuous line),
PSA at positivity threshold (black dashed line),
Redfield (blach dot-dashed line).
In all the Panels the Bose-Einstein occupation number $n_b(\omega_0):=\frac{1}{e^{\beta \omega_0}-1}$ is indicated by a gray thin full line. 
$n_b(\omega_0)$ is the steady state value $\langle a^\dag a\rangle(\infty)$ in the SA. 
For all the Panels we chose the following values of the master equation parameters:
$k_{\rm B} T= 0.5 \omega_0$, $\kappa_0=0.1$,  $\omega_c=5 \omega_0$.
\label{fig:qho-momenta}}
\end{figure}
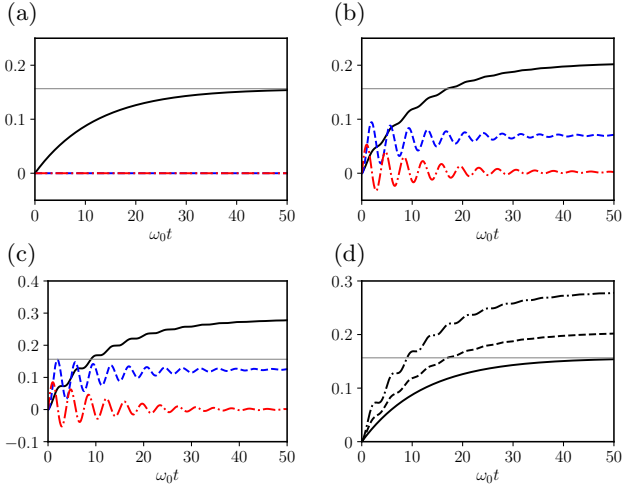

Here the system operator $\zeta$ is the annihilation operator $a\,.$
The Lamb shift acquires from the PSA counter-rotating, squeezing terms $a^2, a^{\dagger 2}\,$
\begin{equation}
H_{\rm LS}^{(\Delta t)}= (\eta_{--} + \eta_{++}) a^\dagger a + \eta_{-+}^{(\Delta t)} a^2 + \eta_{+-}^{(\Delta t)} a^{\dagger 2} + \eta_{--}\,,\\
\end{equation}
implying the non-commutation 
\begin{equation}
\left[H_{\rm S}\,,\, H_{\rm LS}^{(\Delta t)}\right]=2 \omega_0 (\eta_{+-}^{(\Delta t)} a^{\dagger 2}-h.c.)\,.
\end{equation}  
The associated full Hamiltonian (\ref{HAMDT})  has hence the form 
\begin{equation}
H_{\rm S}^{(\Delta t)}=
\bar{\omega} a^\dagger a + \eta_{-+}^{(\Delta t)} a^2 + \eta_{+-}^{(\Delta t)} a^{\dagger 2}\,,
\end{equation}
with
$\bar{\omega}:=\omega_0+\eta_{--}+\eta_{++}$, that, due to the presence of the squeezing terms, 
will induce a non-trivial internal dynamics for S. 
\begin{widetext}
Here the dissipator reads as follows:
\begin{eqnarray}
\mathcal{D}^{(\Delta t)}[\rho_{\rm S}(t)]=
\gamma_{--} \left(a^\dagger \rho_{\rm S}(t) a 
- \frac{1}{2} \left\{a a^\dagger  \,,\, \rho_{\rm S}(t)\right\} \right)
+ 
\gamma_{++} \left(a \rho_{\rm S}(t) a^\dagger 
- \frac{1}{2} \left\{a^\dagger a  \,,\, \rho_{\rm S}(t)\right\} \right)\nonumber\\
+
\gamma_{- +} ^{(\Delta t)}
\left( a \rho_{\rm S}(t) a - \frac{1}{2} \left\{a^2  \,,\, \rho_{\rm S}(t)\right\} \right)
+
\gamma_{+ -} ^{(\Delta t)} 
\left(a^\dagger \rho_{\rm S}(t) a^\dagger - \frac{1}{2} \left\{{a^\dagger}^2  \,,\, \rho_{\rm S}(t)\right\}
\right)\;,
\, \nonumber 
\end{eqnarray}
and
the PSA implies again a breaking of commutation rules at the level of super-operators.
Specifically, as anticipated in Eqs.(\ref{Hs-D-commutator-PSA}) and (\ref{H-D-commutator-PSA}),
both the bare and the full  Hamiltonians don't commute with the dissipator:
\begin{eqnarray}
\label{Hs-D-commutator-PSA-qho}
\left[\mathcal{D}^{(\Delta t)} , \mathcal{H}_{\rm S}\right]_-(\rho)=
- i \omega_0
\left[
\gamma_{-+}^{(\Delta t)}
\left(	2 a \rho  a - a^2 \rho - \rho a^2	\right)-{\rm h.c.}
\right]\,,
\end{eqnarray}
\begin{eqnarray}
\label{H-D-commutator-PSA-qho}
\left[\mathcal{D}^{(\Delta t)} , \mathcal{H}_{\rm S}^{(\Delta t)} \right]_-
(\rho)
=
- i 
\left\{
\left[
\bar{\omega} \gamma_{-+}^{(\Delta t)} - \eta_{-+}^{(\Delta t)} (\gamma_{++} + \gamma_{--})
\right]
\left(	2 a \rho  a - a^2 \rho - \rho a^2	\right)-{\rm h.c.}
\right\}
\nonumber \\
-2 i
\left[
\gamma_{-+}^{(\Delta t)}  \eta_{+-}^{(\Delta t)}  
\left(	a^\dagger \rho a + a \rho a^\dagger - a^\dagger a \rho - \rho a^\dagger a - \rho	\right)
-{\rm h.c. } 
\right]
\,.
\end{eqnarray}
About the effects on the dynamics, the presence of squeezing introduces a coupling between the second momenta of the system: 
\begin{eqnarray}
%
%
\dot{\langle a^2 \rangle}&=&-2i\left(\bar{\omega}\langle  a^2\rangle + 2 \eta_{+-}^{(\Delta t)} 
\langle a^\dag a\rangle+\eta_{+-}^{(\Delta t)}\right)- \gamma_{+-}^{(\Delta t)}-(\gamma_{++}-\gamma_{--})\langle a^2\rangle\,, \\
\dot{\langle a^\dag a \rangle}&=&2i\left(\eta_{-+}^{(\Delta t)} \langle a^2 \rangle - \eta_{+-}^{(\Delta t)} \langle{a}^2 \rangle^* \right)-(\gamma_{++}-\gamma_{--})\langle a^\dag a\rangle  + \gamma_{--}\,.
\end{eqnarray}
\end{widetext}
In Fig.~\ref{fig:qho-momenta} we plot 
$\langle a^\dag a (t)\rangle$, ${\rm Re}[ \langle a^2(t) \rangle]$, ${\rm Im} \langle a^2(t) \rangle$ choosing a bosonic bath ($q=1$)
and considering as initial state the ground state  $\rho_{\rm S}(0)=\ket{0}\bra{0}$ of $H_{\rm S}\,.$
From the Figure we observe that PSA, when compared with SA, introduces oscillations at short time scales and --- contrarily to the example of qubit --- modifies the steady state of the system as well.
\section{Sufficient conditions for the coarse graining time}\label{SEC.iv}  
Going beyond the examples discussed in Sec.~\ref{sec:partial}, 
we are now interested in determining  general conditions which guarantee that 
 a certain finite coarse graining time $\Delta t$ can be adopted to enforce a proper PSA, 
  ensuring that the matrix $\gamma^{(\Delta t)}$ of elements 
 $\gamma^{(\Delta t)}_{ij}$ defined in Eq.~(\ref{DEFGAMMA})
is positive semi-definite, i.e. $\gamma^{(\Delta t)}\geq 0$.
Formally speaking this consists in finding the values of $\Delta t$ such that 
 \begin{equation} \label{posi} 
\vec{u}^\dag \cdot   \gamma^{(\Delta t)} \cdot \vec{u}=  
\sum_{ij} u_i^* \gamma^{(\Delta t)}_{ij} u_j\geq 0\;,
\end{equation}
for all choices of 
the column vector $\vec{u} \in \mathbb{C}^{{N}}$ or, equivalently, 
such that given the minimum eigenvalue $\Lambda_{\min}(\Delta t)$ of $\gamma^{(\Delta t)}$ is non-negative, i.e. 
\begin{eqnarray}
\Lambda_{\min}(\Delta t) \geq 0\;.\label{cond1} \end{eqnarray}
For small values of $N$,  Eq.~(\ref{cond1}) turns out to be the proper way to go as explicitly
verified in the previous section.
 However, as $N$ increases determining $\Lambda_{\min}(\Delta t)$ can be problematic. In what follows we hence present
an alternative, computational less demanding, approach which allows one
to charaterize the set of suitable $\Delta t$, by  only focusing on the properties of the 
$M\times M$ blocks  $\Omega(\omega)$ defined in Eq.~(\ref{MATRIXOMEGA}).
The main result of this analysis is the identification of a
critical threshold $\Delta  t_c$ above which the coarse graining time $\Delta t$ is
guaranteed to yield a positive semi-definite  $\gamma^{(\Delta t)}$, i.e. 
 \begin{eqnarray}  \label{SUFF1} 
\Delta t \geq \Delta t_c \quad \Longrightarrow \quad \gamma^{(\Delta t)}\geq 0\;.
\end{eqnarray} 
Specifically 
indicating
with $\| \Omega(\omega) \|_{\infty}$ the operator norm of $\Omega(\omega)$, i.e. 
 \begin{eqnarray}
\| \Omega(\omega) \|_{\infty} := \sup_{\vec{v}(\omega)} \tfrac{\sqrt{ | \vec{v}^\dagger (\omega) \cdot \Omega^\dag(\omega)\Omega(\omega) \cdot \vec{v}(\omega)|}}{|\vec{v}(\omega)|} \;, 
\end{eqnarray} 
and with 
$\lambda_{\min}(\omega)$ the minimum eigenvalue of its Hermitian component 
 $\gamma^{(+)}(\omega,\omega)$ defined in Eq.~(\ref{DEFGAMMAOMEGAOMEGA})
  (which is non-negative by construction),  
 in the next subsection we shall proof that one can identify $\Delta t_c$  with the quantity 
\begin{eqnarray} \label{DELTACDEF} 
 \Delta t_c^{(1)} &:=& 2 (G-1)  \max_{\omega, \omega': \omega\neq \omega'} \left(
\tfrac{ \| \Omega(\omega) \|_{\infty} + \| \Omega(\omega') \|_{\infty}%
}{
|\omega -\omega'| \; \lambda_{\rm min}(\omega) 
}\right), 
\end{eqnarray} 
or with its pejorative, but more compact, version 
\begin{eqnarray} \label{DELTACDEFsecond} 
 \Delta t_c^{(2)}&:=&  
\frac{ 4 (G-1)\|\Omega\|_{\max}}{\nu_{\min}\; \lambda_{\min}} \;, 
\end{eqnarray} 
where $\lambda_{\min}:= \min_\omega  \lambda_{\rm min}(\omega)$,
$\|\Omega\|_{\max}:= \max_\omega \| \Omega(\omega') \|_{\infty}$, 
and where
\begin{eqnarray}  \label{DEFNU} 
\nu_{\min}:= \min_{\omega, \omega': \omega\neq \omega'}|\omega -\omega'|\;,\end{eqnarray}  is minimum among all the gaps differences.
As $\Delta t_c^{(2)}$  is always larger than $\Delta t_c^{(1)}$, it provides a
worst estimation of the real critical threshold $\Delta t_c$. Still Eq.~(\ref{DELTACDEFsecond}) is more informative as it makes explicit that $\Delta t_c$ should scales as the inverse of the minimal difference $
\nu_{\min}$.
An estimation of the critical time $\Delta t_c$  that is 
provably better, but more involved than
 $\Delta t_c^{(1)}$ is finally given by the quantity
 \begin{eqnarray} \label{DELTACDEF1} 
 \Delta t_c^{(0)}&:=&   \max_{\omega} \left(\frac{2}{{Q(\omega) K(\omega)}{\lambda_{\min}(\omega)}}\right) \;,
\end{eqnarray} 
obtained by the functions 
\begin{eqnarray} \label{qui}
Q(\omega) &:=& \sum_{\omega':\,\omega'\neq \omega} \tfrac{ |\omega-\omega'|}{
 \| \Omega(\omega) \|_{\infty} + \| \Omega(\omega') \|_{\infty}} \;, \\ \label{quo}
 q_{\omega'}^{(\omega)}&:=&  \tfrac{ |\omega-\omega'|}{
 \| \Omega(\omega) \|_{\infty} + \| \Omega(\omega') \|_{\infty}} \tfrac{1}{Q(\omega)}\;,  \quad  (\forall\omega'\neq \omega)\;, \\
 K{(\omega)}&:=&   \frac{1}{\sum_{\omega':\,\omega'\neq \omega} \frac{1}{q^{(\omega)}_{\omega'}}}\;. \label{qua} 
\end{eqnarray}

\subsection{Derivation of the bounds via matrix dilution} 
Here we explicitly show that both the terms Eq.~(\ref{DELTACDEF}) and (\ref{DELTACDEF1}) are suitable choices for the critical time $\Delta t_c$ entering Eq.~(\ref{SUFF1}). 

We start by observing that by expanding the indexes $i$ and $j$, Eq.~(\ref{posi}) can be conveniently casted in the following 
form  
\begin{eqnarray}
\label{eq:same-and-different-omega}
&&\sum_{\omega} \vec{u}^\dagger (\omega) \cdot \gamma^{\rm (+)} (\omega, \omega)
\cdot \vec{u}(\omega)
\\ \nonumber
&& +\sum_{\omega,\omega':\, \omega \neq \omega'} S_{\omega - \omega'}^{(\Delta t)} \; \vec{u}^\dagger (\omega) \cdot \gamma^{\rm (+)} (\omega, \omega')\cdot \vec{u}(\omega')
\geq 0 \,,
\end{eqnarray}
where 
for given $\omega$ and $\omega'$, 
\begin{eqnarray} \label{defgamma} 
\gamma^{\rm (+)} (\omega, \omega')  := 
\Omega^\dag(\omega) +  \Omega(\omega') \;,\end{eqnarray} 
represents the $M\times M$ matrix with elements provided 
 by the terms $\gamma^{(+)}_{\alpha \omega,\beta \omega'}$ of  Eq.~(\ref{gamma-pm-def}), and where 
 $\vec{u}(\omega)$ 
is the $M$-dimensional vector defined by 
the components of 
$\vec{u}$  associated with the corresponding block $\omega$, i.e. $\vec{u}= (\vec{u}({\omega_1}), ..., \vec{u}(\omega_G))^T$.

It is worth observing that the first contribution
of Eq.~(\ref{eq:same-and-different-omega}) corresponds to the term one would get when enforcing secular
approximation (i.e. enforcing the $\Delta t\rightarrow \infty$ limit): accordingly, for all choices of $\vec{u}$ this term can always
be guaranteed to be non negative, i.e.   
\begin{eqnarray}
\label{eq:same-and-different-omega1}
&&\sum_{\omega} \vec{u}^\dagger (\omega) \cdot \gamma^{\rm (+)} (\omega, \omega)
\cdot \vec{u}(\omega)
\geq 0 \,.
\end{eqnarray}
Problems on the contrary can arise from the second contribution which involves the off-diagonal blocks 
$\gamma^{\rm (+)} (\omega, \omega')$ with $\omega\neq \omega'$. 
To treat them we adopt the following {\it dilution} technique dividing 
 the contribution coming from the diagonal block term   $\omega=\omega'$ into fractions which are then added 
to the terms associated with the  off-diagonal blocks $\omega\neq\omega'$.
Specifically, for each given $\omega$  let us introduce a set of numbers  $\{ p_{\omega'}^{\rm (\omega)}\}_{\omega'}$  such that 
\begin{equation}
\label{eq:probability-distribution}
\begin{cases}
p_{\omega'}^{\rm (\omega)}\geq 0\,,\,\omega'\neq \omega\;, \\
\sum_{\omega': \,\omega'\neq \omega}p_{\omega'}^{\rm (\omega)}=1\,.
\end{cases}
\end{equation}
They form $G$  sets  of probabilities with $G-1$ entries, which we shall employ as  free parameters in our analysis 
and which allow us to rewrite (\ref{eq:same-and-different-omega}) in the following symmetrized form 
\begin{widetext} 
\begin{eqnarray}
\label{eq:pos-cond-with-prob}
\sum_{\omega,\omega':\,\omega'> \omega} 
\Big\{\,	
p_{\omega'}^{\rm (\omega)}\;  \vec{u}^\dagger (\omega) \cdot \gamma^{\rm (+)} (\omega, \omega)\cdot \vec{u}(\omega) &+& p_{\omega}^{\rm (\omega')}\;  \vec{u}^\dagger (\omega') \cdot \gamma^{\rm (+)} (\omega', \omega')\cdot \vec{u}(\omega') \nonumber \\
&+& \; 2\; S_{\omega - \omega'}^{(\Delta t)} \; \mbox{Re} \left[ \vec{u}^\dagger (\omega) \cdot \gamma^{\rm (+)} (\omega, \omega')\cdot\vec{u}(\omega') \right]
\,\Big\}
\geq 0\,,
\end{eqnarray} 
where we grouped together all the contributions of all the couples $\omega$ and $\omega'\neq \omega$, and used the fact that $S_{\omega - \omega'}^{(\Delta t)}$ is invariant under exchange of $\omega$ and $\omega'$, and the identity $\gamma^{\rm (+)} (\omega', \omega)=[\gamma^{\rm (+)} (\omega, \omega')]^\dag$.

Now a sufficient condition ensuring that Eq.~(\ref{eq:pos-cond-with-prob}) holds for all $\vec{u}$, can be
obtained by forcing each one of such contributions to verify the same property. 
More specifically, we can claim that the matrix $\gamma^{(\Delta t)} $ is non-negative
at least for those $\Delta t$ such that, there exists
a proper choice of the probabilities $\{ p_{\omega'}^{\rm (\omega)}\}_{\omega'}$ for which 
\begin{eqnarray}\label{eq:cond-fixed-omega-omega1}	
{\cal F}_{\omega,\omega'}^{(\Delta t)}(\vec{u}(\omega),\vec{u}(\omega')):= 
p_{\omega'}^{\rm (\omega)}\;  \vec{u}^\dagger (\omega) \cdot \gamma^{\rm (+)} (\omega, \omega)\cdot \vec{u}(\omega) &+& p_{\omega}^{\rm (\omega')}\;  \vec{u}^\dagger (\omega') \cdot \gamma^{\rm (+)} (\omega', \omega')\cdot \vec{u}(\omega')\nonumber \\
&+& 2\; S_{\omega - \omega'}^{(\Delta t)} \; \mbox{Re} \left[ \vec{u}^\dagger (\omega) \cdot \gamma^{\rm (+)} (\omega, \omega')\cdot\vec{u}(\omega') \right]\geq 0\;,
\end{eqnarray} 
for all possible choices of $\omega$, $\omega'$, $\vec{u}(\omega)$, and $\vec{u}(\omega')$.
Next step is  to construct a lower  bound for the quantity ${\cal F}_{\omega,\omega'}^{(\Delta t)}(\vec{u}(\omega),\vec{u}(\omega'))$.
For this purpose we begin observing that, indicating with $\lambda_{\min}(\omega)$ the minimum eigenvalue of
the matrix $\gamma^{\rm (+)} (\omega, \omega)$, we have
\begin{eqnarray} 
\vec{u}^\dagger (\omega) \cdot \gamma^{\rm (+)} (\omega, \omega)\cdot \vec{u}(\omega)
\geq  |\vec{u}({\omega})|^2 \; \lambda_{\min}(\omega) \;, \label{LOWER1} 
\end{eqnarray} 
with  $|\vec{u}({\omega})|$ being the norm of the vector 
$\vec{u}({\omega})$.
Then by using 
Eq.~(\ref{defgamma}), 
the Cauchy-Schwartz inequality, and  the fact that for generic $\vec{u}$ one has
${\sqrt{| \vec{u}^\dagger (\omega) \cdot \Omega^\dag(\omega)\Omega(\omega) \cdot \vec{u}(\omega)|}}
 \leq   |\vec{u}({\omega})| \; \| \Omega(\omega) \|_{\infty}$, we observe that
\begin{eqnarray}
\label{eq:cauchy-schwartz-ineq}
\Big| {\rm Re}\left[ \vec{u}^\dagger (\omega) \cdot \gamma^{\rm (+)} (\omega, \omega')\cdot \vec{u}(\omega')
\right]\Big|
&\leq& | \vec{u}^\dagger (\omega) \cdot \gamma^{\rm (+)} (\omega, \omega')\cdot \vec{u}(\omega')|
\leq 
 | \vec{u}^\dagger (\omega) \cdot \Omega^\dag(\omega) \cdot \vec{u}(\omega')| +
 | \vec{u}^\dagger (\omega') \cdot \Omega(\omega')\cdot \vec{u}(\omega)|\nonumber \\
 &\leq&
  |\vec{u}({\omega'})|\;  \sqrt{ 
 | \vec{u}^\dagger (\omega) \cdot \Omega^\dag(\omega) \Omega(\omega) \cdot \vec{u}(\omega)|} + 
  |\vec{u}({\omega})| \; \sqrt{ | \vec{u}^\dagger (\omega') \cdot \Omega^\dag(\omega')\Omega(\omega') \cdot \vec{u}(\omega')|} \nonumber \\
  &\leq&  |\vec{u}({\omega'})|  |\vec{u}({\omega})|\left(\| \Omega(\omega)\|_\infty +
  \| \Omega(\omega')\|_\infty \right) \;, 
   \end{eqnarray}
which implies
\begin{eqnarray}
 2\; S_{\omega - \omega'}^{(\Delta t)} \; \mbox{Re} \left[ \vec{u}^\dagger (\omega) \cdot \gamma^{\rm (+)} (\omega, \omega')\cdot\vec{u}(\omega') \right] &\geq&
-  2\; |S_{\omega - \omega'}^{(\Delta t)} |\; \Big| \mbox{Re} \left[ \vec{u}^\dagger (\omega) \cdot \gamma^{\rm (+)} (\omega, \omega')\cdot\vec{u}(\omega') \right]\Big|  \nonumber \\ 
&\geq&  - 2 |S_{\omega - \omega'}^{(\Delta t)}| 
 |\vec{u}({\omega'})|  |\vec{u}({\omega})|\left(\| \Omega(\omega)\|_\infty +
  \| \Omega(\omega')\|_\infty \right)\;. \label{LOWER2} 
 \end{eqnarray} 
 Replacing hence~(\ref{LOWER1})  and (\ref{LOWER2}) into the definition of 
 ${\cal F}_{\omega,\omega'}^{(\Delta t)}(\vec{u}(\omega),\vec{u}(\omega'))$  we arrive to establish the following bound
 \begin{eqnarray}  \label{LOWER3} 
{\cal F}_{\omega,\omega'}^{(\Delta t)}(\vec{u}(\omega),\vec{u}(\omega')) \geq 
{\bar{\cal F}}_{\omega,\omega'}^{(\Delta t)}(\vec{u}(\omega),\vec{u}(\omega'))\;,
\end{eqnarray}
with ${\bar{\cal F}}_{\omega,\omega'}^{(\Delta t)}(\vec{u}(\omega),\vec{u}(\omega'))$ being the function\begin{eqnarray} 
{\bar{\cal F}}_{\omega,\omega'}^{(\Delta t)}(\vec{u}(\omega),\vec{u}(\omega'))&:=& 
 p_{\omega'}^{\rm (\omega)}\; |\vec{u}({\omega})|^2 \; \lambda_{\min}(\omega) + p_{\omega}^{\rm (\omega')}\;  |\vec{u}({\omega'})|^2 \; \lambda_{\min}(\omega')   - 2 |S_{\omega - \omega'}^{(\Delta t)}| 
 |\vec{u}({\omega'})|  |\vec{u}({\omega})|\left(\| \Omega(\omega)\|_\infty +
  \| \Omega(\omega')\|_\infty \right)\nonumber  \\ \label{lambda-equation}
&=& |\vec{u}({\omega})|^2 (A_{\omega, \omega'}-B^{(\Delta t)}_{\omega, \omega'})+|\vec{u}({\omega'})|^2(A_{\omega', \omega}-B^{(\Delta t)}_{\omega, \omega'})+B^{(\Delta t)}_{\omega, \omega'}(|\vec{u}({\omega})|-|\vec{u}({\omega'})|)^2\;, \label{LOWER4} 
 \end{eqnarray} 
\end{widetext} 
with
\begin{eqnarray}
\label{A-and-B-defA}
A_{\omega, \omega'}&:=&p_{\omega'}^{\rm (\omega)} 
\lambda_{\rm min}(\omega)\;,\\
\label{A-and-B-defB}
B^{(\Delta t)}_{\omega, \omega'}&:=&|S_{\omega - \omega'}^{(\Delta t)}| \, 
 \left(\| \Omega(\omega)\|_\infty +
  \| \Omega(\omega')\|_\infty \right)\;, 
\end{eqnarray}
From Eq.~(\ref{LOWER3}) 
it then follows that a sufficient condition for Eq.~(\ref{eq:cond-fixed-omega-omega1}) is the positivity
of the function ${\bar{\cal F}}_{\omega,\omega'}^{(\Delta t)}(\vec{u}(\omega),\vec{u}(\omega'))$, which 
by looking at (\ref{LOWER4}), 
 can be guaranteed by imposing the function $B^{(\Delta t)}_{\omega, \omega'}$  to be smaller than 
$A_{\omega, \omega'}$ and $A_{\omega', \omega}$, i.e. 
\begin{eqnarray}
B^{(\Delta t)}_{\omega, \omega'} \leq \min\{ A_{\omega, \omega'},A_{\omega', \omega}\} \;,
\end{eqnarray} 
which can be casted in the 
 equivalent form 
\begin{eqnarray}
\label{eq:sinc-upper-bound-norm-and-eigen}
|S_{\omega - \omega'}^{(\Delta t)}|\leq
\frac{p_{\omega'}^{\rm (\omega)}
 \lambda_{\rm min}(\omega)}{\| \Omega(\omega)\|_\infty + \| \Omega(\omega')\|_\infty}
\,,
\end{eqnarray} 
by exploiting  the symmetry $B^{(\Delta t)}_{\omega, \omega'}= B^{(\Delta t)}_{\omega', \omega}$. 
Noticing that from Eq.~(\ref{eq:S-function-def}) 
we have
$|S_{\omega - \omega'}^{(\Delta t)}| \leq 2/ (|(\omega-\omega')| \Delta t)$,
the latter can then be replaced by the (stronger) requirement 
\begin{eqnarray}
\label{eq:Delta-t-lower-bound}
\Delta t \geq 
\frac{2}{p_{\omega'}^{\rm (\omega)}} \; 
\frac{
{\| \Omega(\omega)\|_\infty + \| \Omega(\omega')\|_\infty}
}{
|\omega-\omega'|
 \; \lambda_{\rm min} (\omega)
} 
\,.
\end{eqnarray}
To summarize, 
 any coarse graining time $\Delta t$ admitting a set of probability functions 
$\{p_{\omega'}^{\rm (\omega)}\}_{\omega,\omega'}$ for which the inequality~(\ref{eq:Delta-t-lower-bound}) holds for all $\omega$ and $\omega'$, ensures the fulfilment of Eq.~(\ref{eq:pos-cond-with-prob}), hence the non-negativity of the matrix $\gamma^{(\Delta t)}$ 
(notice that if $\lambda_{\rm min} (\omega)=0$ for some $\omega$, Eq.~(\ref{eq:Delta-t-lower-bound}) can still be used: simply it implies that $\Delta t$ has to be infinite). 
Alternatively, we can say that for each assigned choice of  the dilution probabilities
(\ref{eq:probability-distribution}) 
the fulfilment of the inequality Eq.~(\ref{eq:pos-cond-with-prob}) allows us to identify a coarse graining
time $\Delta t$ that implies the non-negativity of $\gamma^{(\Delta t)}$.
Taking for instance $\{p_{\omega'}^{\rm (\omega)}\}_{\omega'}$ to be  flat distributions, i.e. 
\begin{eqnarray}
p_{\omega'}^{\rm (\omega)} = 1/(G-1)\;, \qquad \forall \omega'\neq \omega\end{eqnarray} 
equation (\ref{eq:Delta-t-lower-bound}) becomes 
\begin{eqnarray}
\label{eq:Delta-t-lower-bound1}
\Delta t \geq 
2 (G-1)
\frac{
{\| \Omega(\omega)\|_\infty + \| \Omega(\omega')\|_\infty}
}{
|\omega-\omega'|
 \; \lambda_{\rm min} (\omega)
} 
\,,
\end{eqnarray}
which maximizing the r.h.s. term with respect to all possible choices of $\omega$ and 
$\omega'\neq \omega$, 
allows us to claim that a sufficient condition for the non-negativity of $\gamma^{(\Delta t)}$
can be obtained by  taking  $\Delta t$ larger than the quantity $\Delta t_c^{(1)}$ 
of Eq.~(\ref{SUFF1}). 

To prove that also Eq.~(\ref{DELTACDEF1}) yields a legittimate  estimation of 
$\Delta t_c$, we look for the optimal choice 
of the probability functions $\{p_{\omega'}^{\rm (\omega)}\}$ entering Eq.~(\ref{eq:Delta-t-lower-bound}).
To see this let us use the functions~(\ref{qui})-(\ref{qua}) to 
rewrite the latter inequality as 
\begin{equation}
\label{eq:Delta-t-lower-boundINV}
\frac{2}{ \lambda_{\min}({\omega}) \Delta t} \leq 
\frac{p_{\omega'}^{(\omega)} |\omega-\omega'|}
{\| \Omega(\omega)\|_\infty + \| \Omega(\omega')\|_\infty}
=Q(\omega) p_{\omega'}^{(\omega)} q_{\omega'}^{(\omega)} \;.
\end{equation}
Now observe that for given $\omega$, similarly to the $\{ p_{\omega'}^{(\omega)}\}_{\omega'}$,
the terms $\{ q_{\omega'}^{(\omega)}\}_{\omega'}$   define a proper
set of probabilities with $G-1$ entries.
As we have the freedom to arbitrarily choose whatever set of $\{ p_{\omega'}^{(\omega)}\}_{\omega'}$, 
in order to get a less stringent condition on $\Delta t\,,$ we want to focus on those that maximize the
r.h.s. of  Eq.~(\ref{eq:Delta-t-lower-boundINV}). A simple proof by contradiction shows that this can be achieved by ensuring that, for all given $\omega$, the quantity
$p_{\omega'}^{(\omega)}  q_{\omega'}^{(\omega)}$ should be constant in $\omega'$, for all $\omega'\neq \omega$. By imposing the normalization condition it then follows that such constant must coincide
with the function $K(\omega)$ defined in Eq.~(\ref{qua}), i.e.
$p_{\omega'}^{(\omega)}  q_{\omega'}^{(\omega)} = K(\omega)$ which inserted into 
Eq.~(\ref{eq:Delta-t-lower-boundINV}) yields 
\begin{equation}
\label{eq:Delta-t-lower-boundINV1}
\frac{2}{\lambda_{\min}({\omega}) \Delta t} \leq 
Q(\omega) K(\omega) \Longleftrightarrow \Delta t \geq \frac{2}{\lambda_{\min}({\omega}) 
Q(\omega) K(\omega)}
\end{equation}
that upon maximization over $\omega$ finally leads to Eq.~(\ref{DELTACDEF1}).

\section{Summary and conclusions}
Starting from the Redfield equation the SA is a standard procedure to ensure completely positive dynamics. 
It is also known that it is equivalent to an infinitely large choice of the coarse grain time scale.
On the contrary the PSA keeps such time scale finite.
Using a general formalism we found sufficient conditions to guarantee the complete positivity of the Redfield equation, among which a tight bound on the coarse grain time interval. Furthermore we explicitly show that non-secular terms can determine non commutation between the Hamiltonian and the dissipating parts of the master equation.
We thus provide examples by applying the partial secular approximation to a qubit or harmonic oscillator interacting with a fermionic or bosonic thermal environment via dipole-like interaction.
\\ 

Acknowledgement: we thank G. M. Andolina, M. Polini, V. Cataudella and G. De Filippis for useful comments.

\appendix 

\section{Positivity of the matrix $\gamma^{\rm (+)}(\omega,\omega)$} \label{appA} 
We discuss here the positivity of the secular blocks $\gamma_{\alpha \omega, \beta \omega}^{\rm (+)}$ of the matrix $\gamma_{\alpha \omega, \beta \omega_1}^{(\Delta t)}$ of Eq.\,(\ref{DEFGAMMA}),
by following the demostration given in \cite{breuer2002theory}. 
Such blocks are the ones with equal frequencies and their entries read as
\begin{equation}
\label{appA:fourier-transform}
\gamma_{\alpha \omega, \beta \omega}^{\rm (+)}
=
\int_{-\infty}^{+\infty} c_{\alpha \beta}(\tau) e^{i \omega \tau} d\tau\,,
\end{equation}
with $c_{\alpha \beta}(\tau)$ being the bath correlation functions given in Eq.~(\ref{eq:bath-correlations}).
We should prove that
\begin{equation}
\label{appA:positivity1}
\sum_{\alpha \beta} u^*_{\alpha}(\omega) \gamma_{\alpha \omega, \beta \omega}^{\rm (+)} u_\beta (\omega)
\geq
0 
\end{equation}
for any  $\vec{u}(\omega) \in \mathbb{C}^M\,.$
The above expression is actually the Fourier transform of a function $f(\tau)\,:$
\begin{equation}
\label{appA:positivity2}
\sum_{\alpha \beta} u^*_{\alpha}(\omega) \gamma_{\alpha \omega, \beta \omega}^{\rm (+)} u_\beta (\omega)=
\int_{-\infty}^{+\infty} e^{i \omega \tau} f(\tau) d\tau 
\end{equation}
with
\begin{equation}
\label{appA:f}
f(\tau):=\langle \Theta^{\dagger}(\tau)\Theta(0)\rangle\,,
\end{equation}
\begin{equation}
\label{appA:theta}
\Theta(\tau):=\sum_\alpha u_\alpha \tilde{B}_\alpha(\tau)\,,
\end{equation}
with $\tilde{B}_\alpha(\tau)$ being the operators on the thermal bath of Eq.~(\ref{eq:bath-correlations}).
From the function $f(\tau)$ it is possible to define an $n\times n$ matrix $f_{lm}$ in the following way:
\begin{equation}
\label{appA:f-matrix}
f_{lm}:=f(\tau_l-\tau_m)=
\langle \Theta^{\dagger}(\tau_l)\Theta(\tau_m)\rangle\,,
\end{equation}
with $\tau_i \in \{\tau_1, \tau_2, ..., \tau_n\}\,.$
Such matrix is positive semi-definite for any choices of the times $\tau_l$ and of the dimension $n$.
This can be proven by using the fact that the trace of the product of two positive semi-definite operators is non negative. In formulas:
\begin{equation}
\label{appA:positivity-flm}
\sum_{l m} v^*_l f_{lm} v_m = \langle \Delta^\dagger \Delta \rangle:={\rm tr}\{ \rho_{\rm E} \Delta^\dagger \Delta\} \geq 0\,,
\end{equation} 
for any complex vectors $\vec{v}\,,$
with
\begin{equation}
\Delta:=\sum_m v_m \Theta(\tau_m) \,.
\end{equation}
From the positivity of the matrix 
$f_{lm}$ it follows that the Fourier transform of $f(\tau)$ is always non-negative (Bochner's theorem) and hence the positivity of the matrix $\gamma_{\alpha \omega, \beta \omega}^{\rm (+)}$ is guaranteed (see Eq.~(\ref{appA:positivity2})).
\\
This can be understood by thinking integrals as summations.
Indeed the Fourier transform in the right-hand side of Eq.~(\ref{appA:positivity2}) can be written in a form which is analogous to the left-hand side of Eq.\,(\ref{appA:positivity-flm}) which we know to be a positive quantity:
\begin{eqnarray}
&&\int_{-\infty}^{+\infty}
ds'
e^{i \omega s'}
f(s')
 \\\nonumber
&&\quad=\frac{1}{2T} \int_{-T}^{+T}dl
 \int_{-\infty}^{+\infty}ds\,
u^*(s) f(s-l) u(l)
\geq
0\,,
\end{eqnarray}
with $
u(\tau):=e^{-i \omega \tau}$
and for any $T\,.$
%
%
%
%

%
%
%
%
%

\end{document}